\documentclass[a4paper,fleqn]{cas-dc}

\usepackage[authoryear]{natbib}
\def\tsc#1{\csdef{#1}{\textsc{\lowercase{#1}}\xspace}}
\tsc{WGM}
\tsc{QE}
\usepackage{booktabs}
\usepackage{tabularx}

\usepackage{multirow}

\begin{document}
\let\WriteBookmarks\relax
\def\floatpagepagefraction{1}
\def\textpagefraction{.001}

\shorttitle{Tail Dependence in EU Carbon Markets}

\shortauthors{Maciejowski and Leonelli}

\title[mode=title]{Tail Dependence in EU Carbon Markets: Graphical Models of Extremes for EUA Futures}

\tnotemark[1]
\tnotetext[1]{This work was supported by the Spanish Agencia Estatal de Investigaci\'{o}n grant PID2023-153222OB-I00.}

\author[1]{Jan Maciejowski}
\ead{j.maciejowski@student.ie.edu}
\credit{Conceptualization, Methodology, Software, Formal analysis, Data curation}

\affiliation[1]{organization={School of Science and Technology, IE University},
            city={Madrid},
            country={Spain}}

\author[1]{Manuele Leonelli}
\cormark[1]
\fnmark[1]
\ead{manuele.leonelli@ie.edu}
\ead[url]{https://manueleleonelli.github.io/}
\credit{Conceptualization, Methodology, Supervision, Validation, Writing}

\cortext[1]{Corresponding author}
\fntext[1]{0000-0002-2562-5192}
\begin{abstract}
Understanding how extreme price movements propagate across financial
and energy markets is critical for risk management and regulatory design
in the EU Emissions Trading System (EU ETS). We apply H\"{u}sler-Reiss
graphical models of extremes to a system of 20 daily variables centred
on EU allowances futures across Phases 3 and 4 of the EU ETS
(2013--2025), with a Gaussian graphical model as the average-dependence
baseline. The tail networks are structurally distinct from the average
dependence network: substantially denser, organized around different
central nodes, and governed by within-sector homophily that binds sector
boundaries more tightly than at the average-dependence level. EU
allowances futures are peripheral in the standard graphical model but
achieve the highest centrality in the tail networks, while equity
indices and major FX pairs follow the opposite trajectory. Exponential
random graph models confirm equity and FX peripherality in tail networks
across all sample periods and identify triadic closure during market
downturns as a Phase~3 phenomenon that vanishes in Phase~4. The phase
transition restructures the tail network without thinning it: average
dependence contracts sharply while tail dependence persists, and crash
contagion shifts from clustered to diffuse propagation. These findings
have direct implications for hedge construction by compliance entities,
stress-test calibration by regulators, and the design of systemic-risk
monitoring tools for EU ETS markets.
\end{abstract}

%\item First application of graphical models of extremes to a carbon market.
%\item EU allowances are peripheral in the average dependence network but central %in the tails.
%\item Phase~4 average dependence weakens sharply while tail dependence persists.
%\item Coal, gas, and oil remain EU allowances' strongest tail partners across both phases.
%\item Crash contagion shifts from clustered to diffuse propagation between Phase~3 and Phase~4.


\maketitle

%%=============================================================
\section{Introduction}\label{sec:intro}
%%=============================================================

The EU Emissions Trading System (EU ETS) is the world's largest carbon
market and a cornerstone of European climate policy. Under the system,
firms subject to emissions obligations must hold European Union Allowances
(EUAs) to cover their output, creating a financial market for carbon
permits whose price dynamics are central to both compliance strategy and
investment decisions. As the ETS has matured through successive regulatory
phases, most recently strengthened under the Fit for 55 package, EUA prices
have become increasingly integrated with energy commodity markets, equity
indices, and macroeconomic indicators. Understanding the structure of
these interdependencies is essential for regulators, compliance entities,
and financial institutions active in the carbon market.

A substantial empirical literature has sought to characterize how EUA
prices co-move with their financial and energy drivers, using a wide range
of econometric and statistical tools including structural vector
autoregression \citep{lovcha2021}, quantile regression \citep{zheng2021},
structural equation modeling and regime-dependent specifications
\citep{wang2020,leitao2021}, copula and copula-GARCH approaches
\citep{berrisch2023,trabelsi2023}, non-parametric information-theoretic
methods \citep{salvagnin2024,salvagnin2025}, and probabilistic graphical
models \citep{maciejowski2026}. While these methods have substantially
advanced understanding of EUA price dynamics under normal market
conditions, the most consequential co-movements occur not on average but
at the extremes: price crashes driven by regulatory announcements,
geopolitical shocks, or energy crises, and price spikes associated with
supply tightening or demand surges. Recent quantile connectedness studies
confirm that total connectedness among EUA and energy markets rises from
approximately 28\% at the median to 77\% at extreme quantiles
\citep{wei2023,cao2024}, indicating that average-dependence methods
substantially understate the systemic linkage of EUA markets precisely
when this linkage matters most.

Existing analyses of EUA extremes are nonetheless severely constrained in
dimension. \citet{feng2012} provide the foundational univariate
application, combining GARCH filtering with peaks-over-threshold
estimation to obtain VaR and expected shortfall for EUA alone. Bivariate
copula extensions \citep{reboredo2015,hanif2021a,trabelsi2023} extend this
to one partner variable at a time but cannot condition on the broader
system, and quantile connectedness frameworks \citep{wei2023,cao2024}
characterize marginal spillovers but cannot recover the joint conditional
extremal dependence structure. The reason these methods stop at bivariate
or marginal characterizations is that multivariate extreme value analysis
in dimension $d > 2$ has long been computationally and statistically
demanding \citep[see, e.g.,][]{de2017,leonelli2020semiparametric,nolan2024modeling}.
A recent line of research has addressed this challenge by introducing
graphical models for extremes
\citep{engelke2020,hentschel2025statistical}, which encode
conditional independence in the joint tail through a graph structure
analogous to Gaussian graphical models (GGMs). Applications of extremal graphical models range from flood risk assessment \citep{asadi2015}
to financial returns \citep{kluppelberg2021,rama2026} and
food nutrition \citep{buck2021}, but the framework has not yet been
brought to bear on carbon markets.

This paper closes that gap. We apply H\"{u}sler-Reiss (HR) graphical models of
extremes to a multivariate system of 20 daily variables centred on EUA
futures, spanning commodity, equity, clean energy, volatility, FX, and
bond categories. We estimate sparse extremal graphs
separately for positive (simultaneous price spikes) and negative
(simultaneous crashes) tails, and compare against a GGM baseline that captures average dependence. Exponential
random graph models are then fitted to each learned network to formally
relate its topology to sector attributes, providing a principled
structural decomposition of tail connectivity. The analysis is conducted
on the full Phase~3--4 sample (2013--2025) as well as on each phase
separately, a split motivated by the regulatory regime change between
Phase~3 and Phase~4 \citep{borghesi2023,dittmann2025}, which coincided
with the COVID-19 shock, the Russo-Ukrainian war, the energy crisis, and
the Fit for 55 reforms.

This is the first application of graphical models of
extremes to any carbon market. Two themes emerge from the analysis.
First, the tail dependence network is structurally distinct from the
average dependence network in a way that reverses the structural role of
nodes: EUA futures, peripheral in the standard graphical model, achieve
the highest centrality in the tail networks, while equity indices and
major FX pairs, dominant under average dependence, become peripheral
under stress. Second, the Phase~3-to-Phase~4 transition restructures the
tail network without thinning it: average dependence contracts sharply
while tail dependence persists, and the mode of crash contagion shifts
from clustered to diffuse propagation, consistent with the broader
financialization of EU ETS markets in Phase~4
\citep{borghesi2023,terranova2025}. These findings have direct
implications for hedge construction by compliance entities, stress-test
calibration by regulators, and the design of systemic-risk monitoring
tools as the EU ETS evolves through further regulatory tightening and
the planned integration of additional sectors.

The paper is organized as follows. Section~\ref{sec:lit} reviews related
literature. Section~\ref{sec:data} describes the data and modeling
framework. Section~\ref{sec:results} reports results.
Section~\ref{sec:discussion} discusses implications and limitations.
Section~\ref{sec:conclusion} concludes.

%%=============================================================
\section{Literature Review}\label{sec:lit}
%%=============================================================

A wide range of academic studies has sought to explain the drivers of European Union
Allowance (EUA) prices across regulatory, commodity, energy, macroeconomic, and
speculative dimensions. This body of work, typically based on standard econometric
techniques, has yielded important insights but varies considerably in its ability to capture
joint extreme co-movements and structural changes in EUA market behavior across trading
phases. We summarize findings across key domains, motivating the variables included in our
extremal graphical model and clarifying how our approach extends the state of the art.

%%-------------------------------------------------------------
\subsection{Regulatory mechanisms}
%%-------------------------------------------------------------

The supply of EUAs is governed by regulatory instruments that fundamentally shape price
dynamics in the EU ETS. Among these, the Market Stability Reserve (MSR) and the Linear
Reduction Factor (LRF) are the most influential. The MSR, introduced in 2019 and
strengthened under the Fit for 55 package, adjusts auction volumes based on the Total
Number of Allowances in Circulation: when this exceeds 833 million allowances, 24\% of the
surplus is withheld; when it falls below 400 million, allowances are released. Under the
revised rules, any holdings in the reserve above the lower threshold are permanently
invalidated, tightening supply on a structural basis. \citet{borghesi2023} provide a
comprehensive review of the MSR, finding that while the mechanism successfully supported
EUA prices during the COVID-19 demand shock by absorbing excess supply, it also shortens
the effective banking horizon and may amplify price volatility in response to unexpected
shocks. The LRF was raised from 2.2\% to 4.3\% under Fit for 55, compressing the cap
trajectory toward the 2030 climate targets.

These supply-side constraints define a structural break between Phase~3 (2013--2020) and
Phase~4 (2021--2025), the two subperiods we study. The transition between phases is not
merely administrative: it corresponds to a qualitative change in the market's regulatory
environment, its participant composition, and its sensitivity to external shocks
\citep{borghesi2023,dittmann2025}. Regulatory uncertainty regarding the future
configuration of these mechanisms acts as a persistent background driver of EUA price
volatility, a feature our sample split is explicitly designed to capture.

%%-------------------------------------------------------------
\subsection{Commodity markets}
%%-------------------------------------------------------------

EUA prices are closely tied to fossil fuel commodity markets, particularly coal and natural
gas. Their relative prices govern the fuel-switching decision of power producers: when
natural gas is expensive relative to coal, utilities shift toward more carbon-intensive
coal-fired generation, raising the demand for allowances; the reverse holds when gas is
cheaper. This mechanism is well-established across multiple studies and ETS phases
\citep{lovcha2021,tan2017,dittmann2025}. \citet{lovcha2021} employ a structural vector
autoregression with frequency-domain decomposition, finding that commodity fundamentals
explain 65--90\% of EUA price variance at business-cycle frequencies, while high-frequency
variation is dominated by speculative activity. At the phase level, \citet{dittmann2025}
document that energy fundamentals explain up to 12\% of EUA price variance in Phase~3 but
below 1\% in Phase~4, reflecting the deepening financialization of carbon markets: a
central finding motivating our phase-split design.

Oil, though less directly tied to power generation, influences EUA prices as a global energy
benchmark, a driver of natural gas contract prices, and a proxy for aggregate economic
conditions. \citet{zheng2021} decompose oil price shocks into supply, demand, and risk
components and apply quantile regression to EUA returns across Phases~1--3. Supply and
demand shocks exert a positive effect on EUA returns, while risk shocks associated with
financial stress are negative; importantly, these effects are amplified in the lower tail of
the EUA distribution, providing an early signal that the joint distribution behaves
differently in market extremes than at the center.

%%-------------------------------------------------------------
\subsection{Energy markets and clean energy}
%%-------------------------------------------------------------

The energy sector is the largest source of verified emissions under the EU ETS, making
developments in electricity generation and renewable energy directly relevant to EUA
pricing. Changes in electricity prices reflect shifts in fuel costs and carbon permit costs
simultaneously, creating two-way feedback between power markets and EUA markets. Equity
indices tracking the renewable energy sector, including the ERIX index and the NYSE
Bloomberg Global WIND Index, have been shown to act as net spillover transmitters to the
EUA market \citep{hanif2021a,maciejowski2026}.

\citet{hanif2021a} apply frequency-domain connectedness measures and bivariate copulas
between EUA returns and six clean energy equity indices over 2011--2020, finding that EUA
consistently receives spillovers from clean energy markets. Notably, they document
asymmetric lower tail dependence between EUA and the WIND Index, the only statistically
significant tail dependence result in the existing bivariate literature involving EUA returns, indicating that EUA and clean energy equities share joint crash risk that is invisible in
linear correlation.

The time-varying nature of EUA-energy dependencies is further documented by
\citet{berrisch2023}, who jointly model EUA, gas, coal, and oil prices using a
VECM-Copula-GARCH framework with time-varying copula parameters estimated on
2010--2022 data. The EUA-gas correlation emerges as the most volatile pairwise
relationship in their system, fluctuating around $+0.3$ before dropping sharply to $-0.4$
in the weeks following the Russian invasion of Ukraine in February 2022. This structural
instability in EUA-energy dependence provides further justification for analyzing the
extremal network separately for Phase~3 and Phase~4.

%%-------------------------------------------------------------
\subsection{Macroeconomic and financial conditions}
%%-------------------------------------------------------------

European equity markets, exchange rates, and bond markets shape EUA pricing through
their effect on expectations about industrial activity and investor sentiment. \citet{wang2020}
apply a BN with structural equation modeling to Phase~3 data, finding that
European equity indices and Brent oil directly affect EUA returns, while broader indices
operate through indirect channels mediated by the stock market. Bond markets convey risk
sentiment via investor positioning: \citet{leitao2021} find that green bond indices exert a
positive and significant effect on EUA prices across high- and low-volatility regimes, while
conventional bonds are negatively associated with EUA prices in volatile markets.

The relative weight of macroeconomic versus commodity drivers has shifted substantially
across phases. \citet{salvagnin2024}, applying the non-parametric Information Imbalance to
data from January 2014 to April 2023, find that Phase~3 EUA prices are primarily driven by
energy variables (coal, gas, and the ERIX renewable index) while Phase~4 is dominated by
financial and currency variables.  A causal extension by \citet{salvagnin2025}, using the
Differentiable Information Imbalance, identifies the IBEX35 and coal futures as the most
causally relevant predictors across the full sample. This shift reflects the financialization of
carbon markets (characterized by the growing role of investment banks, hedge funds, and
algorithmic traders) which is widely recognized as a defining feature of Phase~4
\citep{borghesi2023,terranova2025}.

%%-------------------------------------------------------------
\subsection{Speculative dynamics and market structure}
%%-------------------------------------------------------------

Speculation plays a limited role in long-run EUA price trends but is a major driver of
short-term variability and the formation of price bubbles \citep{lovcha2021,terranova2025}.
The EUA market exhibits structural characteristics conducive to speculative activity:
trading is highly concentrated, with 97\% of secondary volume occurring on the
Intercontinental Exchange (ICE), and the participant base has evolved toward financial
institutions. By 2023, the European Securities and Markets Authority (ESMA) reported that
73\% of secondary market volume involved financial intermediaries, a share that reflects the
broader financialization of Phase~4 markets.

\citet{terranova2025} apply the bubble-detection procedure of \citet{phillipsshi2020}
to EUA
futures and options data from 2017 to 2022, identifying seven speculative bubbles present
on approximately 10\% of trading days, with six of the seven triggered by regulatory
announcements rather than commodity fundamentals. Quantile connectedness studies
reinforce this picture: \citet{wei2023} find that EUA is a consistent net receiver of risk
spillovers across all market conditions, while \citet{cao2024} document that total
connectedness among EUA and energy markets rises from approximately 28\% at the median
to 77\% at extreme quantiles. This divergence between median and extreme connectedness
implies that standard average-dependence methods substantially understate the systemic
linkage of EUA markets during periods of tail stress: precisely the regime our paper is
designed to characterize.

%%-------------------------------------------------------------
\subsection{Modeling landscape}
%%-------------------------------------------------------------

Studies of EUA price dynamics have employed a wide range of methods that differ in their
treatment of nonlinearity, regime changes, and joint distributional behavior. SVAR and
frequency-domain frameworks trace mean responses across markets \citep{lovcha2021};
quantile regression and Markov-switching models examine tail heterogeneity and regime
dependence \citep{tan2017,leitao2021,zheng2021}; structural equation modeling and BNs map conditional dependence in the center of the distribution
\citep{wang2020,maciejowski2026}; and copula-based approaches, including DCC-GARCH
variants and VECM-Copula-GARCH, focus on time-varying pairwise dependence
\citep{berrisch2023,trabelsi2023}. Non-parametric predictive methods
\citep{salvagnin2024,salvagnin2025} improve variable-selection performance but are designed
for forecasting rather than for recovering joint distributional structure. Quantile
connectedness frameworks \citep{wei2023,cao2024} characterize extreme spillovers on
marginal distributions but cannot recover the conditional extremal independence structure
of the joint tail.

Within the extreme value literature, \citet{feng2012} provide the foundational application
to EUA markets, combining GARCH filtering with peaks-over-threshold estimation of the
generalized Pareto distribution to obtain univariate VaR and expected shortfall for EUA
prices across Phases~1 and~2. They document that downside risk exceeds upside risk and
explicitly call for future work extending the univariate EVT framework to a multivariate
setting -- precisely the direction pursued here. \citet{reboredo2015} take a partial step in
this direction, fitting bivariate EGARCH-EVT-copula models to EUA-oil and EUA-gas pairs
using Phase~2 data; their finding that the Gaussian copula best fits both pairs, implying no
tail dependence between EUA and fossil fuels in Phase~2, provides a baseline that our
analysis updates for the Phase~3--4 environment. More recently, \citet{fang2021}
apply semiparametric GAS models to estimate time-varying extreme risk for EUA and Chinese
carbon allowances across Phase~3, identifying two structural breaks in EUA tail risk at June
2014 and March 2017. \citet{trabelsi2023} extend the bivariate copula approach to
time-varying parameter specifications between Certified Emission
Reductions and energy market variables, finding that copulas with tail
dependence parameters outperform static alternatives and that
speculative activity drives the strength of tail coupling during oil
market disturbances. A more recent line of research, initiated by
\citet{engelke2020}, has developed multivariate graphical models for
extremes, in which conditional independence in the joint tail is
encoded by the graph structure of an HR Pareto distribution.

%%-------------------------------------------------------------
\subsection{Summary and outlook}
%%-------------------------------------------------------------

The literature reviewed above documents the drivers of average EUA price variation and the
structural shift between Phase~3 and Phase~4 across multiple methodological frameworks
\citep{borghesi2023,salvagnin2024,dittmann2025}. There is growing evidence that EUA
market behavior under stress diverges from its average behavior: extreme connectedness
substantially exceeds median connectedness \citep{wei2023,cao2024}, and bivariate analyses
find tail dependence between EUA and clean energy equity indices that linear correlation
does not capture \citep{hanif2021a}. Yet no existing study recovers the joint conditional
extremal dependence structure of EUA futures and its related markets as a whole, and none
applies graphical models of extremes to any carbon market.

The present paper addresses this gap. Building on the EVT foundations of \citet{feng2012}
and the graphical extremes framework of \citet{engelke2020}, we estimate HR
graphical models of the joint tail distribution of EUA futures together with a broad set of
commodity, energy, and financial variables, separately for Phase~3 and Phase~4. Comparing
the resulting extremal networks with Gaussian graphical models of average dependence
reveals where the tails diverge from the center, and an ERGM-based interpretation links
network topology to sector structure. The result is a characterization of EUA extreme
co-movement risk that is complementary to, but fundamentally different from, what the
existing literature can offer.

%%=============================================================

\section{Materials and Methods}

\subsection{Data}\label{sec:data}

We assemble a set of
20 daily variables covering the commodity, energy, equity, macroeconomic, and
financial dimensions of EUA price formation. The variable set, described in
Table~\ref{tab:variables} and identical to that used in \citet{maciejowski2026}, spans
Phases~3 and~4 of the EU ETS from January 3, 2013 to January 31, 2025, sourced from
Bloomberg. It covers ICE EUA futures (MO1), energy commodities (Brent crude oil, API2
Rotterdam coal, natural gas), major European and US equity indices (STOXX Europe 600,
CAC 40, DAX, S\&P 500), the MSCI Europe Energy Index, clean energy equity indices
(S\&P Global Clean Energy, WilderHill ECO), volatility measures (VIX, gold spot), five
Euro exchange rate pairs (USD, GBP, CHF, CNY, RUB), and two bond indices (Bloomberg
Pan-European High Yield, Bloomberg EuroAgg Investment Grade). For the purpose of
network analysis, variables are assigned to six sectors: Commodity, Equity, Clean
Equity, FX, Volatility, and Bond. The variable set is designed to span the dimensions identified by the literature as primary drivers of EUA pricing: energy commodities for the fuel-switching channel; equity indices for industrial-demand and risk-sentiment channels; clean-energy indices for the renewables-supply channel; FX rates for international transmission; volatility and bond indices for risk premium and financial conditions.

\begin{table*}
\centering
\footnotesize
\resizebox{\textwidth}{!}{
\begin{tabularx}{\textwidth}{l l X l}
\toprule
\textbf{Category} & & \textbf{Variable Name} & \textbf{Bloomberg Ticker} \\
\midrule
Dependent Variable & & ICE European Union Allowance Futures (1st month) & MO1 Comdty \\
\midrule
Energy Commodities & & ICE Brent Crude Oil Futures (1st month) & CO1 Comdty \\
 & & ICE API2 Rotterdam Coal Futures (1st month) & XA1 Comdty \\
 & & Natural Gas Futures (1st month, composite pricing) & NG1 COMB Comdty \\
\midrule
Stock Indexes & & S\&P 500 Index & SPX Index \\
 & & STOXX Europe 600 Price Index & SXXP Index \\
 & & French CAC 40 & CAC Index \\
 & & German DAX & DAX Index \\
\midrule
Volatility Measures & & Chicago Board Options Exchange Volatility Index & VIX Index \\
 & & Gold Spot Rate & XAU Curncy \\
\midrule
Energy Indexes & & MSCI Europe Energy Index & MXEU0EN Index \\
 & & S\&P Global Clean Energy Index & SPGTCED Index \\
 & & WilderHill Clean Energy Index & ECO Index \\
\midrule
Bond Indexes & & Bloomberg Pan-European High Yield Total Return Index & LP01TREU Index \\
 & & Bloomberg EuroAgg Total Return Index Value Unhedged EUR & LBEATREU Index \\
\midrule
FX Rates & & Euro to U.S. Dollar Currency Exchange Rate & EURUSD Curncy \\
 & & Euro to British Pound Sterling Currency Exchange Rate & EURGBP Curncy \\
 & & Euro to Swiss Franc Currency Exchange Rate & EURCHF Curncy \\
 & & Euro to Chinese Renminbi Currency Exchange Rate & EURCNY Curncy \\
 & & Euro to Russian Ruble Currency Exchange Rate & EURRUB Curncy \\
\bottomrule
\end{tabularx}}
\caption{List of variables included in the analysis, grouped by category, with their corresponding Bloomberg tickers.}
\label{tab:variables}
\end{table*}

Following \citet{mcneil2000}, all series are pre-filtered with AR--GARCH models in
\citet{maciejowski2026} to remove serial autocorrelation and conditional
heteroskedasticity, and the analyses here are conducted on the resulting standardized
residuals $\mathbf{Z} = (z_{i,t})$, where $z_{i,t}$ denotes the residual of variable
$i$ at time $t$. Missing values arising from market holidays are carried forward from
the most recent available observation.

For the standard GGM baseline, the full matrix $\mathbf{Z}$ is used directly. For the
extreme value analysis, each residual series is further transformed to standard Pareto
margins via
\begin{equation}
    x_{i,t} = \frac{1}{1 - \hat{F}_i(z_{i,t})},
\end{equation}
where $\hat{F}_i$ denotes the empirical marginal distribution of $z_{i,t}$. Only
observations exceeding a threshold corresponding to the upper $p = 0.20$ fraction of
the data are retained, yielding the extreme sample $\mathbf{X}$. This transformation
places all series on a common Pareto scale and ensures that the joint tail behaviour
is not confounded by differences in marginal distributions \citep{engelke2020}. To
study negative extremes, simultaneous market crashes, the same procedure is
applied to $-z_{i,t}$, so that the lower tail of returns becomes the upper tail of
the transformed series. All analyses are conducted separately on the full sample,
Phase~3 (January 2013 -- December 2020), and Phase~4 (January 2021 -- January 2025).

\subsection{Standard Gaussian graphical model}

As a baseline characterising average dependence, we estimate a Gaussian
graphical model (GGM). Suppose the standardised residual vector
$\mathbf{Z} = (Z_1, \ldots, Z_d)$ follows a multivariate Gaussian
distribution with precision matrix $\Theta = \Sigma^{-1}$, where
$\Sigma$ is the covariance matrix. The off-diagonal entries of $\Theta$
encode the partial correlation between $Z_i$ and $Z_j$ controlling for
all the remaining variables,
\begin{equation}
\rho_{ij}
\;=\; -\,\Theta_{ij} \,/\, \sqrt{\Theta_{ii}\,\Theta_{jj}}.
\end{equation}
A zero partial correlation corresponds to conditional independence
between $Z_i$ and $Z_j$ given the rest of the system, equivalently to
the absence of an edge between $i$ and $j$ in the graph. The GGM
therefore turns the precision matrix into a network in which edges
represent the conditional co-movement structure that survives once
indirect dependence through other variables has been accounted for.

We adopt a Bayesian estimation approach: we recover the posterior
distribution of the precision matrix under a non-informative prior and
include an edge $(i,j)$ in the graph whenever the 95\% credible
interval of the corresponding partial correlation excludes zero. The
resulting graph is weighted and signed: positive edges indicate
conditional co-movement, negative ones indicate conditional opposition.
Bayesian estimation avoids the need for cross-validation penalty
selection and provides direct uncertainty quantification on edge
inclusion.

\subsection{H\"{u}sler-Reiss graphical model for extremes}

Standard graphical models characterise conditional dependence at the
centre of the joint distribution and can fail entirely to represent how
variables co-move at the extremes. A natural pairwise measure of
extremal dependence is the extremal correlation,
\begin{equation}
\chi_{ij} \;=\; \lim_{u \to 1}
P\!\left(F_j(X_j) > u \mid F_i(X_i) > u\right)
\;\in\; [0, 1],
\end{equation}
which represents the probability that one variable exceeds an extreme
quantile given that another does. A value of $\chi_{ij} = 0$ indicates
asymptotic independence: even when both variables are extreme on their
own, they do not tend to be extreme jointly. A value of $\chi_{ij} > 0$
indicates asymptotic dependence: extreme events tend to co-occur.
Because $\chi_{ij}$ is defined entirely through the joint tail, it can
take very different values from the ordinary Pearson or partial
correlation, and two variables with strong ordinary correlation can be
asymptotically independent.

The HR Pareto distribution plays a role in extremal
dependence modelling analogous to that of the multivariate Gaussian
distribution in classical dependence modelling. Whereas the Gaussian
distribution describes the centre of a multivariate distribution
through a covariance matrix, the HR distribution describes its joint
tail through a symmetric matrix $\Gamma \in \mathbb{R}^{d \times d}$,
called the \emph{variogram matrix}. Entries of $\Gamma$ measure the
strength of pairwise extremal dependence between variables: small
values of $\Gamma_{ij}$ indicate strong tail co-movement, while large
values indicate near-independence in the tail. The HR distribution is
the only multivariate extreme value distribution whose conditional
independence structure can be encoded through a graph in a way directly
analogous to GGMs 
\citep{engelke2020,hentschel2025statistical}.

The associated extremal graphical model defines an undirected graph
$\mathcal{G} = (V, E)$ on the set $V = \{1, \ldots, d\}$ of variables.
An edge between $i$ and $j$ is present in $\mathcal{G}$ if and only if
$X_i$ and $X_j$ are conditionally tail-dependent given all the
remaining variables. The absence of an edge therefore indicates that
the extremal dependence between $X_i$ and $X_j$ is fully mediated by
the other variables in the system, the tail analogue of the conditional
independence interpretation that underpins the GGM.

Estimation is performed on the extreme sample $\mathbf{X}$ defined in
Section~\ref{sec:data}, with positive and negative extremes treated as
separate networks throughout to capture the possibility that
simultaneous spikes and simultaneous crashes propagate through
structurally different conditional dependence networks. The choice of
threshold $p = 0.20$ balances the bias-variance trade-off in tail
estimation: a lower threshold contaminates the sample with non-extreme
observations, while a higher threshold reduces the effective sample
size beyond what the phase-specific analyses can sustain.

The variogram matrix $\Gamma$ and the graph $\mathcal{G}$ are recovered
simultaneously by minimising a penalised log-likelihood,
\begin{equation}
\hat{\Gamma},\, \hat{\mathcal{G}} \;=\;
\arg\min_{\Gamma,\,\mathcal{G}}
\left\{ -\ell(\Gamma; \mathbf{X}) + \rho\,\|\Gamma\|_1 \right\},
\end{equation}
where $\ell$ is the HR log-likelihood, $\|\Gamma\|_1$ is the
off-diagonal $L^1$ norm of $\Gamma$, and $\rho > 0$ is a sparsity
penalty. This is the extremal analogue of the graphical lasso
\citep{friedman2008}: the $L^1$ penalty shrinks small entries of
$\Gamma$ to zero, and the resulting zero pattern defines the estimated
graph $\hat{\mathcal{G}}$. We select $\rho$ by 10-fold cross-validation
maximising held-out log-likelihood. As a complementary summary of the
\emph{unconditional} pairwise tail behaviour of each pair of
variables, we also report the empirical extremal correlation matrix
$\hat{\chi}$ for each sample period and tail direction.

\subsection{Network analysis}

For each learned graph we compute a battery of graph-theoretic
measures that summarise its global topology and the structural role of
individual nodes \citep{newman2018networks}. At the global level we
report the number of edges and the corresponding edge density (the
fraction of possible edges that are present); the diameter (the
longest shortest path between any pair of nodes) and the average
shortest path length, both summarising how compact the network is; the
global transitivity or clustering coefficient (the fraction of
connected triples that close into triangles, indicating whether
dependence propagates in tightly knit clusters); the modularity (the
extent to which the network decomposes into well-separated
communities); the number of communities recovered; and the degree
assortativity coefficient (positive when high-degree nodes
preferentially connect to other high-degree nodes, indicating a
core-periphery pattern, and negative when high-degree nodes preferentially connect to low-degree nodes, indicating a
hub-and-spoke pattern).

At the node level we report four standard centrality measures.
\emph{Degree} counts the number of direct neighbours and gives the
simplest measure of local prominence. \emph{Betweenness} centrality
measures the extent to which a node lies on shortest paths between
other nodes, capturing its role as a structural bridge.
\emph{Eigenvector} centrality weights connections by the centrality of
the neighbour itself, so that a node is central if it connects to
other central nodes; it captures structural prominence rather than
local connectivity. \emph{PageRank} centrality is a related random-walk-based variant: it identifies nodes that are
visited most often by a random walker on the graph.
Community structure is recovered through the Louvain algorithm, which
iteratively assigns nodes to communities so as to maximise the
modularity of the resulting partition.

\subsection{Exponential random graph models}

Graph-theoretic measures describe what a network looks like, but they
do not separate the contributions of distinct structural drivers of
connectivity. To formally relate the topology of each learned network
to observable variable attributes, we estimate exponential random
graph models (ERGMs) \citep{lusher2013exponential}. ERGMs specify the
probability of an observed graph $G$ as
\begin{equation}
\mathbb{P}(G = g) \;\propto\;
\exp\!\left\{\boldsymbol{\theta}^\top \mathbf{s}(g)\right\},
\end{equation}
where $\mathbf{s}(g)$ is a vector of network statistics computed on $g$
and $\boldsymbol{\theta}$ is the corresponding parameter vector. ERGMs
play a role for networks analogous to that of regression for
individual outcomes: the coefficient $\theta_k$ measures the
contribution of network feature $s_k$ to the log-odds of observing the
realised graph, controlling for the other features in the
specification. With $d = 20$
variables, ERGMs are used here as a descriptive inferential device for structural decomposition rather than as a generative model of network formation; we report coefficients with appropriate caution and emphasise within-paper consistency of effects across the nine networks rather than absolute magnitudes.

We estimate a nested sequence of three model specifications. The null
model M0 contains only an edges term, equivalent to an
Erd\H{o}s-R\'{e}nyi random graph in which each pair of nodes is connected
independently with the same probability. The model M1 adds
sector-level nodefactor terms capturing the tendency of each sector to
form connections relative to a reference category (Bond). The model
M2 extends M1 with two further terms: a within-sector homophily term
capturing the tendency of variables within the same sector to be
connected, and a triadic closure term (specifically, the geometrically
weighted edgewise shared partner statistic) capturing the tendency of
connected nodes to share common neighbours, thereby forming triangles.
Model selection across M0--M2 uses the Akaike and Bayesian
information criteria. The sector classification used in the nodefactor and nodematch terms
is the one introduced in Section~\ref{sec:data}, with each variable
assigned to one of Commodity, Equity, Clean Equity, FX, Volatility, or
Bond. EU allowances futures (MO1) are classified as Commodity,
consistent with their role as a traded emissions permit.

All computations were performed in R. Code and data
implementing the methods described above, together with scripts to
reproduce all results and figures in this paper, are available at
\url{https://github.com/manueleleonelli/extremes\_EUA}.

%%=============================================================
\section{Results}\label{sec:results}
%%=============================================================

We present results in five subsections, organized around the paper's primary
contribution: characterizing the tail dependence structure of EUA futures markets.
We first describe the extreme dependence networks for positive and negative tails
(Section~\ref{sec:res:extreme}), then examine how these networks evolve across
EU~ETS phases (Section~\ref{sec:res:phase}). Section~\ref{sec:res:chi} reports
pairwise extremal dependence and EUA's position within it. We then contrast the
extreme networks with the standard GGM baseline
(Section~\ref{sec:res:divergence}), and close with an ERGM decomposition of the
structural drivers of tail connectivity (Section~\ref{sec:res:ergm}). Comprehensive
numerical results (node-level centrality measures, the MO1 row of the
$\hat{\chi}$ matrix, and the complete sequence of ERGM specifications) are
provided in the Appendix.

%%-------------------------------------------------------------
\subsection{Extreme dependence networks}\label{sec:res:extreme}
%%-------------------------------------------------------------

\begin{figure*}[htbp]
    \centering
    \includegraphics[width=0.5\textwidth]{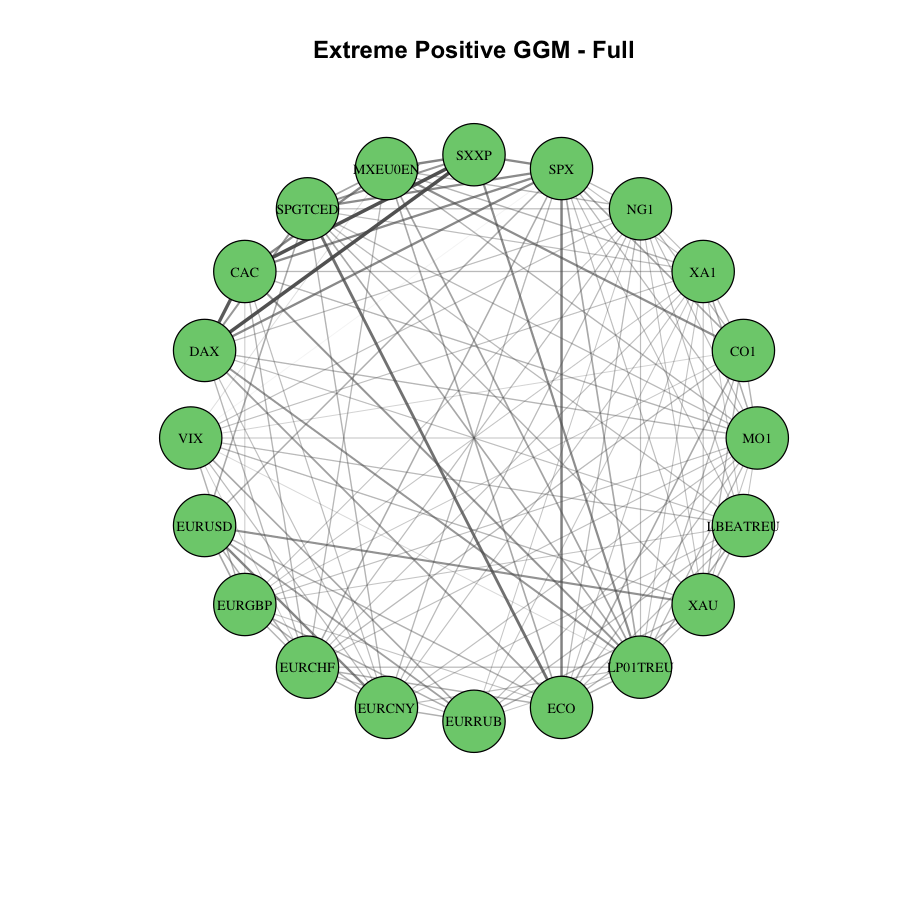}\hfill
    \includegraphics[width=0.5\textwidth]{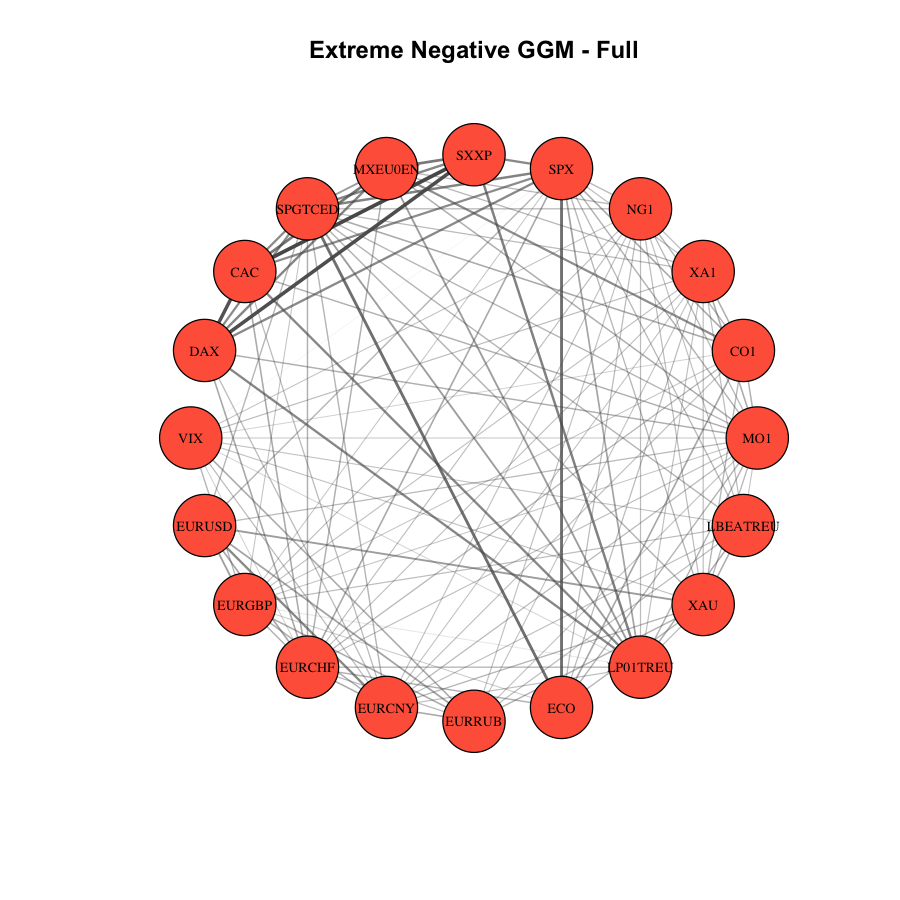}\\\vspace{-1.1cm}
    \includegraphics[width=0.5\textwidth]{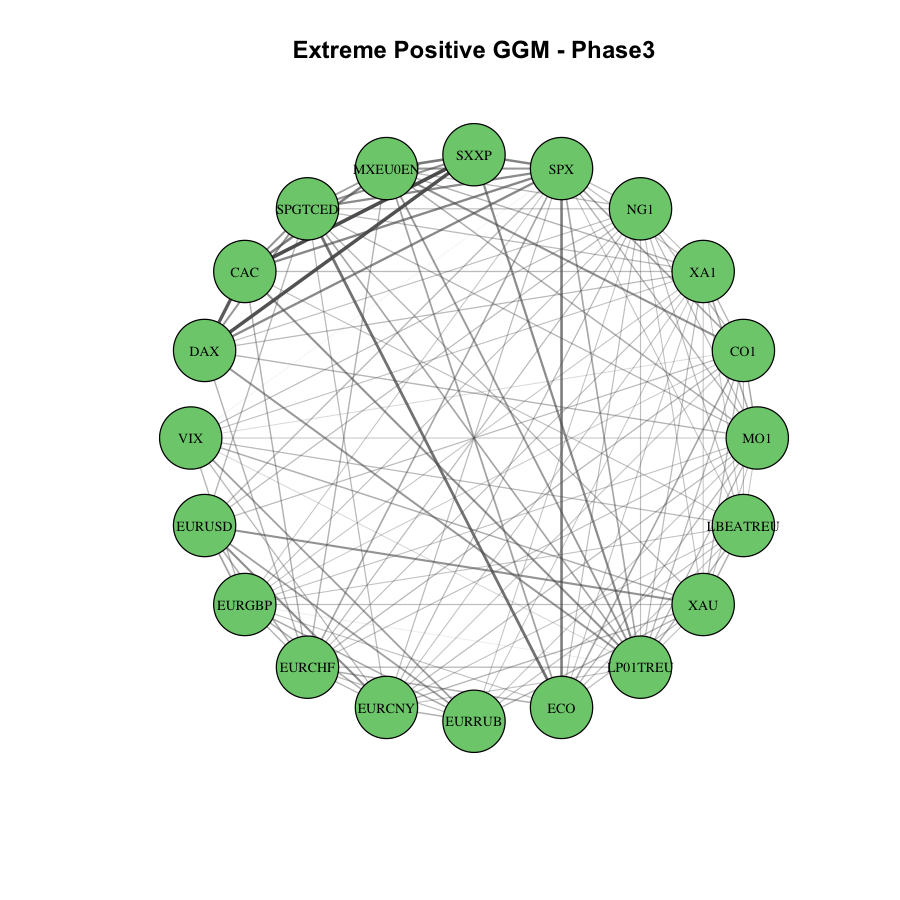}\hfill
    \includegraphics[width=0.5\textwidth]{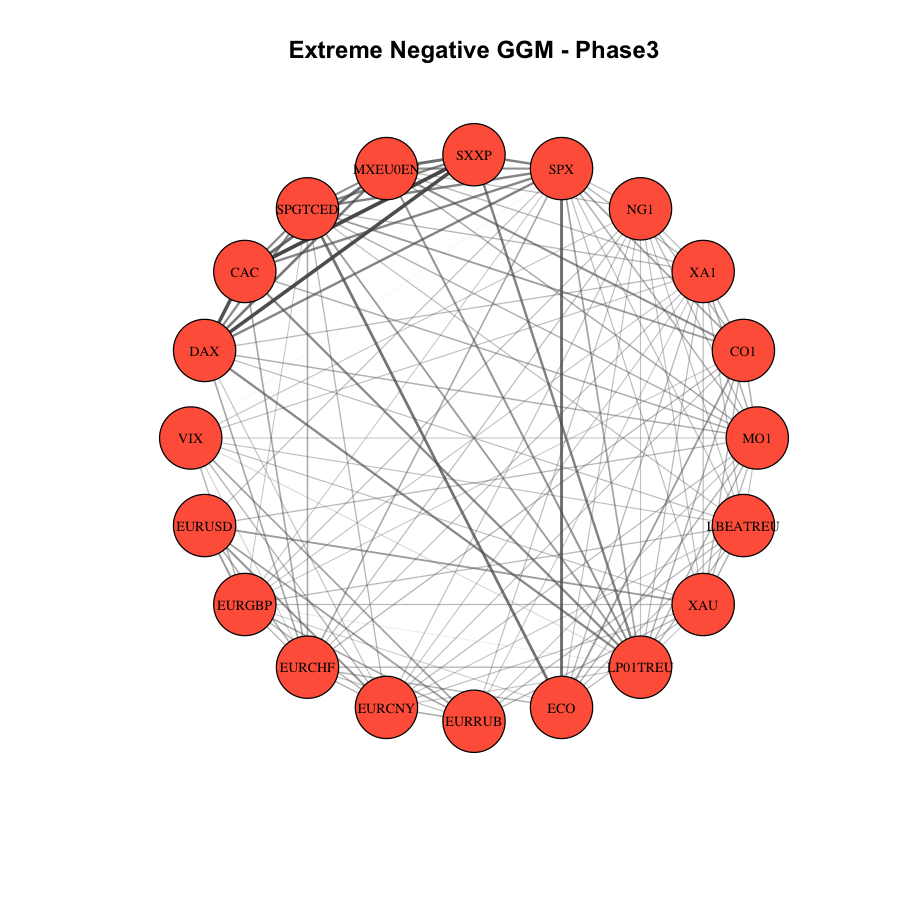}\\ \vspace{-1.1cm}
    \includegraphics[width=0.5\textwidth]{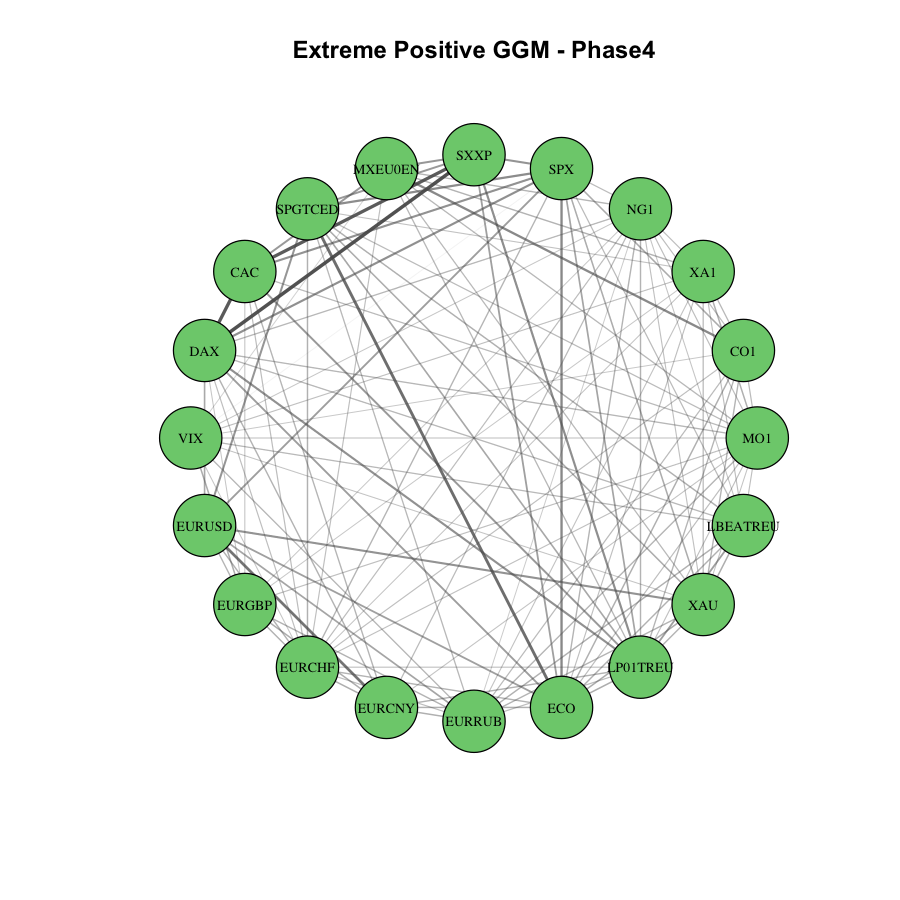}\hfill
    \includegraphics[width=0.5\textwidth]{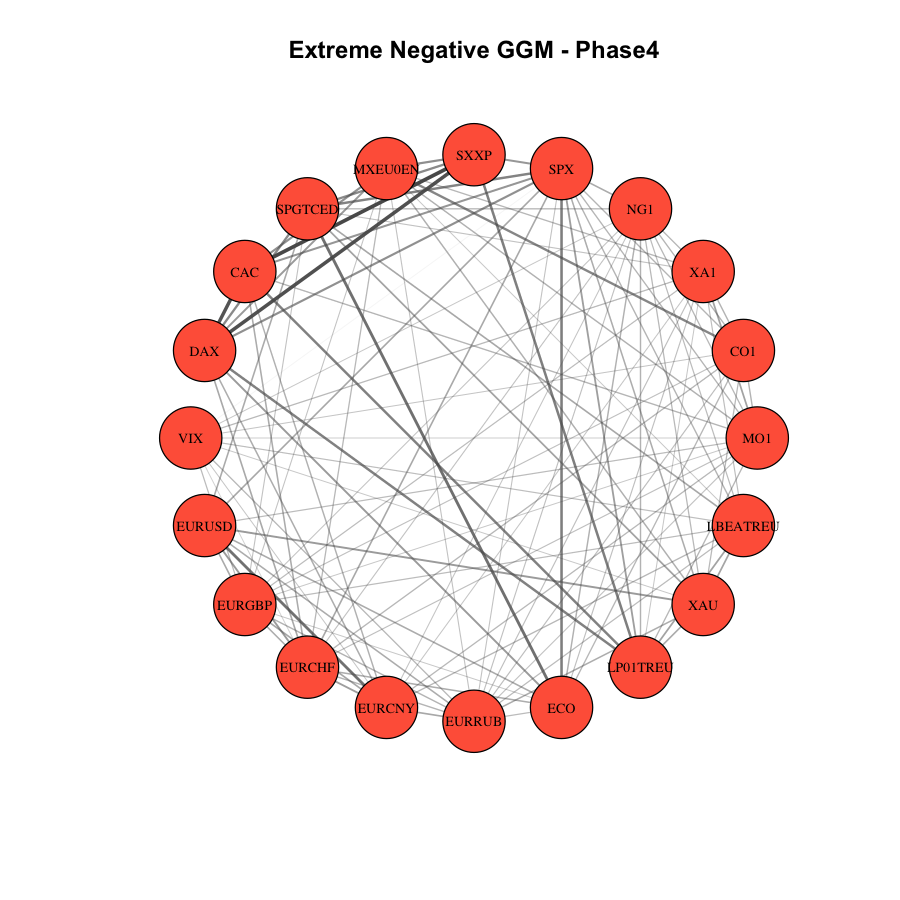}\vspace{-1.1cm}
    \caption{HR extreme graphical models. Left column: positive
    extremes. Right column: negative extremes. Rows: full sample (top), Phase~3 (middle),
    Phase~4 (bottom). Edge width is proportional to extremal dependence strength.}
    \label{fig:extreme}
\end{figure*}

Figure~\ref{fig:extreme} displays the HR graphical models for positive extremes (simultaneous price spikes, left column) and negative extremes
(simultaneous market crashes, right column) across the full sample, Phase~3, and
Phase~4. Table~\ref{tab:globalstats} summarizes global network statistics for all nine estimated networks: three sample periods (full, Phase~3, Phase~4) by three network types (standard GGM, positive extreme, negative extreme).

\begin{table*}[htbp]
\centering
\caption{Global network statistics for all estimated networks across sample
periods. Standard refers to the Bayesian GGM; Pos.\ and Neg.\ extreme refer to the
HR extremal graphical models for the upper and lower tails
respectively.}\label{tab:globalstats}
\resizebox{\textwidth}{!}{
\begin{tabular}{ll rrrrrrrrr}
\toprule
Period & Network & Edges & Density & Avg.\ deg. & Diameter &
Avg.\ path & Transitivity & Modularity & Communities & Assortativity \\
\midrule
\multirow{3}{*}{Full}
  & Standard      & 81  & 0.426 & 8.1  & 3 & 1.61 & 0.500 & 0.155 & 4 & $-$0.102 \\
  & Pos.\ extreme & 128 & 0.674 & 12.8 & 2 & 1.33 & 0.727 & 0.070 & 3 & $-$0.059 \\
  & Neg.\ extreme & 125 & 0.658 & 12.5 & 3 & 1.35 & 0.737 & 0.105 & 2 & $-$0.158 \\
\midrule
\multirow{3}{*}{Phase~3}
  & Standard      & 66  & 0.347 & 6.6  & 3 & 1.78 & 0.468 & 0.195 & 5 & $-$0.073 \\
  & Pos.\ extreme & 127 & 0.668 & 12.7 & 3 & 1.34 & 0.745 & 0.115 & 2 & $-$0.079 \\
  & Neg.\ extreme & 123 & 0.647 & 12.3 & 3 & 1.36 & 0.723 & 0.119 & 2 & $-$0.126 \\
\midrule
\multirow{3}{*}{Phase~4}
  & Standard      & 59  & 0.311 & 5.9  & 3 & 1.81 & 0.374 & 0.186 & 4 & $-$0.053 \\
  & Pos.\ extreme & 119 & 0.626 & 11.9 & 2 & 1.37 & 0.669 & 0.117 & 3 & $-$0.068 \\
  & Neg.\ extreme & 110 & 0.579 & 11.0 & 2 & 1.42 & 0.617 & 0.116 & 3 & \phantom{$-$}0.039 \\
\bottomrule
\end{tabular}}
\end{table*}

The full-sample extreme networks are dense and highly transitive. The
positive extreme network contains 128 edges (density 0.674, average
degree 12.8, transitivity 0.727), while the negative extreme network is
marginally sparser with 125 edges (density 0.658, average degree 12.5,
transitivity 0.737). For context, 190 edges is the maximum for 20
nodes, so the positive extreme network retains approximately
two-thirds of all possible conditional tail dependencies even after
cross-validated penalisation. Such density is expected rather than
anomalous: extreme co-movements in financial systems are known to be
markedly more pervasive than average co-movements, and the
cross-validation criterion selects the sparsity penalty to maximise
held-out log-likelihood rather than to enforce sparsity for its own
sake. The structural conclusions that follow do not rest on density
alone but on which nodes are central, which sectors are peripheral,
and how the topology shifts between phases. Both extreme networks
also have low modularity (0.070 and 0.105), indicating that
conditional tail dependence is diffuse across the system rather than
tightly compartmentalized within sectors.

Table~\ref{tab:mo1} reports the centrality of EUA futures (MO1) across all nine
networks, alongside that of the STOXX Europe 600 (SXXP) for comparison. In the
full-sample extreme networks, MO1 has degree 17 (positive) and degree 18 (negative)
out of a maximum of 19, and achieves the highest eigenvector centrality (1.000) in
both tail directions. It ranks among the top five nodes by eigenvector centrality
in every extreme network configuration, and first in three of six. The variables
most consistently connected to MO1 in the tails span all sectors: commodity (CO1,
XA1, NG1), equity and clean equity (SPX, MXEU0EN, SPGTCED), volatility (VIX, XAU),
FX (EURCHF, EURCNY, EURRUB), and bonds (LP01TREU). SXXP, by contrast, has degree
6--7 throughout the extreme networks and eigenvector centrality between 0.32 and
0.42: placing it among the least connected nodes despite being one of the most
central nodes in the standard GGM.

\begin{table}[htbp]
\centering
\caption{Degree and eigenvector centrality of MO1 (EUA futures) and SXXP (STOXX
Europe 600) across all network types and sample periods, illustrating the
reversal of structural roles between average and tail dependence.}\label{tab:mo1}
\scalebox{0.85}{
\begin{tabular}{ll rr rr rr}
\toprule
 & & \multicolumn{2}{c}{Standard} & \multicolumn{2}{c}{Pos.\ extreme} &
   \multicolumn{2}{c}{Neg.\ extreme} \\
\cmidrule(lr){3-4}\cmidrule(lr){5-6}\cmidrule(lr){7-8}
Node & Period & Deg & Eigen & Deg & Eigen & Deg & Eigen \\
\midrule
\multirow{3}{*}{MO1}
  & Full    & 8  & 0.55 & 17 & 1.00 & 18 & 1.00 \\
  & Phase~3 & 3  & 0.12 & 16 & 0.96 & 16 & 0.94 \\
  & Phase~4 & 3  & 0.24 & 16 & 1.00 & 13 & 0.91 \\
\midrule
\multirow{3}{*}{SXXP}
  & Full    & 12 & 0.86 & 6  & 0.34 & 6  & 0.32 \\
  & Phase~3 & 12 & 0.95 & 6  & 0.32 & 6  & 0.35 \\
  & Phase~4 & 8  & 0.74 & 7  & 0.42 & 6  & 0.34 \\
\bottomrule
\end{tabular}}
\end{table}

The positive and negative full-sample extreme networks share a common core but
differ in community structure. Louvain community detection partitions the
positive extreme network into three groups: a commodity-volatility core (MO1,
CO1, XA1, NG1, MXEU0EN, VIX, ECO, XAU, LBEATREU), an equity-bond cluster
(SPX, SXXP, SPGTCED, CAC, DAX, LP01TREU), and an FX cluster (EURUSD, EURGBP,
EURCHF, EURCNY, EURRUB). The negative extreme network collapses to two
communities: a large commodity-FX-volatility block (MO1, CO1, XA1, NG1, SPX,
VIX, EURUSD, EURGBP, EURCNY, EURRUB, ECO, XAU, LBEATREU) and an equity-bond
block (SXXP, MXEU0EN, SPGTCED, CAC, DAX, EURCHF, LP01TREU). Thus, while sector
boundaries partially organize the positive tail structure, during simultaneous
crashes the FX and commodity nodes merge into a single block and the
distinction between sectors blurs.

The negative extreme network also exhibits stronger degree disassortativity
($-0.158$ versus $-0.059$ for positive), meaning high-degree nodes connect
preferentially to low-degree nodes during crashes, consistent with a
hub-and-spoke propagation pattern in the negative tail.

%%-------------------------------------------------------------
\subsection{Phase evolution of extreme dependence}\label{sec:res:phase}
%%-------------------------------------------------------------

Both extreme networks thin from Phase~3 to Phase~4, but the contraction is
moderate relative to the standard GGM (Table~\ref{tab:globalstats}). Negative
extreme density falls from 0.647 to 0.579 (a loss of 13 edges), while positive
extreme density declines from 0.668 to 0.626 (a loss of 8 edges). Global
transitivity also declines, from 0.723 to 0.617 in the negative tail and from
0.745 to 0.669 in the positive tail. Both directions thus show a moderate
loosening of tail connectivity in the post-2020 environment, though neither
network approaches the sparsity of the standard GGM in either phase.

The phase transition affects MO1 asymmetrically across tail directions. In
positive extremes, MO1 maintains degree 16 in both phases (a near-complete
tail neighborhood) and retains the highest eigenvector centrality (1.000) in
Phase~4 (Table~\ref{tab:mo1}). In negative extremes, however, MO1's degree
drops from 16 in Phase~3 to 13 in Phase~4: five nodes disconnect (DAX, ECO,
LBEATREU, SPGTCED, and XAU), while two new nodes appear (EURCHF and EURGBP).
The contraction is concentrated among equity, clean energy, bond, and volatility
nodes, while commodity connections persist and the FX neighborhood expands.
EUA's crash-tail neighborhood therefore restructures in Phase~4, becoming more
commodity- and FX-focused and losing cross-sector connections to financial
variables.

SXXP's peripherality in the tails is stable across phases: it maintains degree
6--7 and eigenvector centrality around 0.32--0.42 across all extreme networks,
regardless of tail direction or sample period. The STOXX Europe 600 index,
which is the broadest European equity index in the system, is consistently
the least connected node in both tails.

Community structure fragments in Phase~4. In the negative tail, Phase~3
produces two communities (an equity-dominated block containing MO1 and a
commodity-FX-volatility block) but Phase~4 splits into three communities as
coal (XA1) separates from other commodities and joins clean energy and certain
FX nodes (SPGTCED, EURUSD, EURGBP, ECO). In the positive tail, a similar
fragmentation occurs: Phase~3 yields two communities (a
commodity-FX-volatility core with MO1, and an equity block) while Phase~4
separates European equity (CAC, DAX) with FX into a distinct third community.
The energy crisis of 2021--2022 thus appears to have disrupted the
within-sector cohesion that organized the Phase~3 tail network, particularly
in the negative direction.

A notable asymmetry between tail directions persists across both phases. In
Phase~3, the negative extreme network has stronger degree disassortativity
($-0.126$) than the positive ($-0.079$), reflecting concentrated crash
propagation through high-degree hub nodes. In Phase~4, however, the negative
extreme network is the only one with mildly positive degree assortativity
($+0.039$): high-degree nodes now connect preferentially to other high-degree
nodes during crashes. This reversal from disassortative to assortative mixing
in the Phase~4 negative tail suggests a structural change in the crash
propagation mechanism from a hub-mediated pattern in Phase~3 to a
core-periphery pattern in Phase~4.

%%-------------------------------------------------------------
\subsection{Pairwise extremal dependence and the position of EUA
futures}\label{sec:res:chi}
%%-------------------------------------------------------------

\begin{figure*}[htbp]
    \centering
    \includegraphics[width=0.48\textwidth]{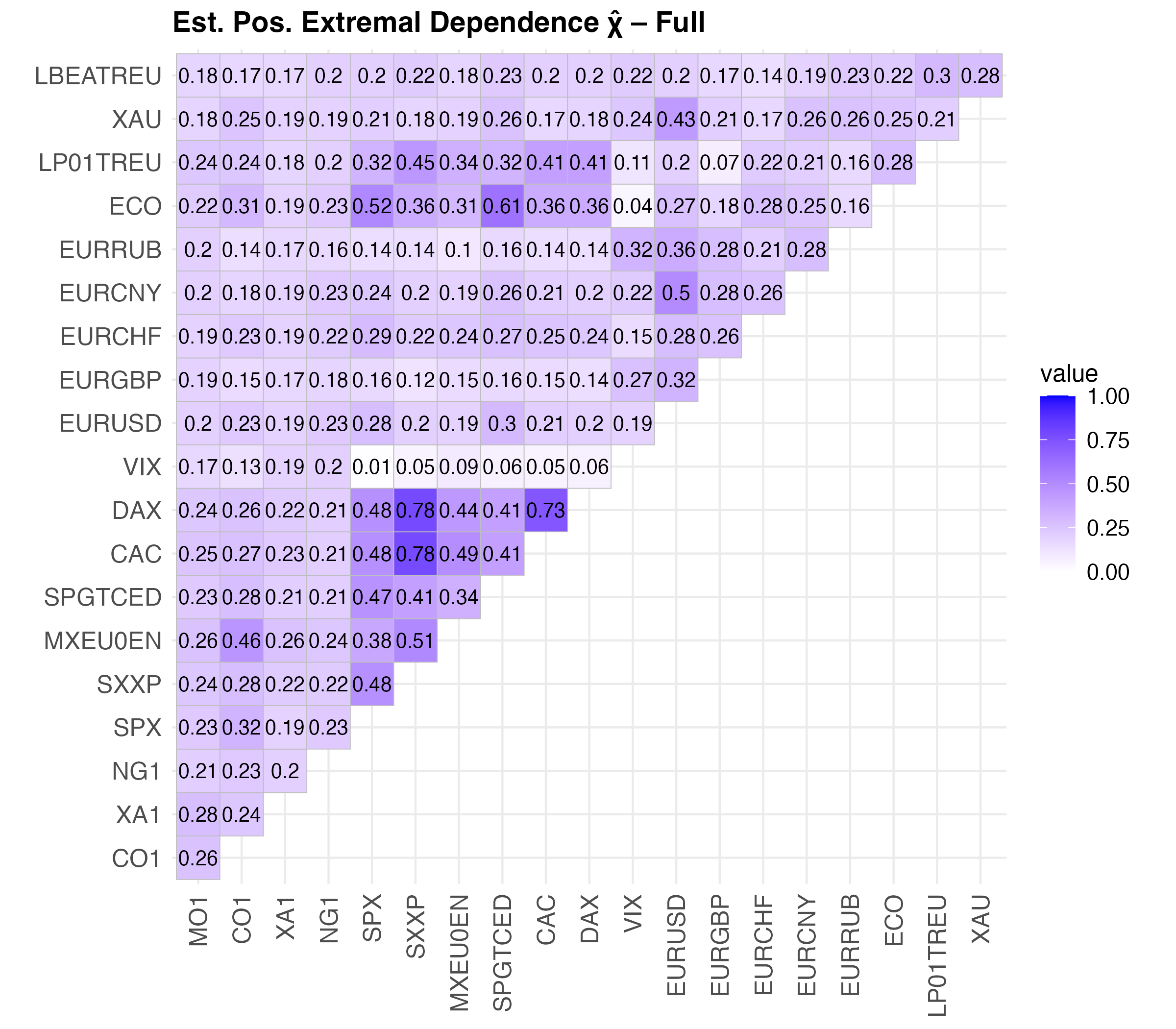}\hfill
    \includegraphics[width=0.48\textwidth]{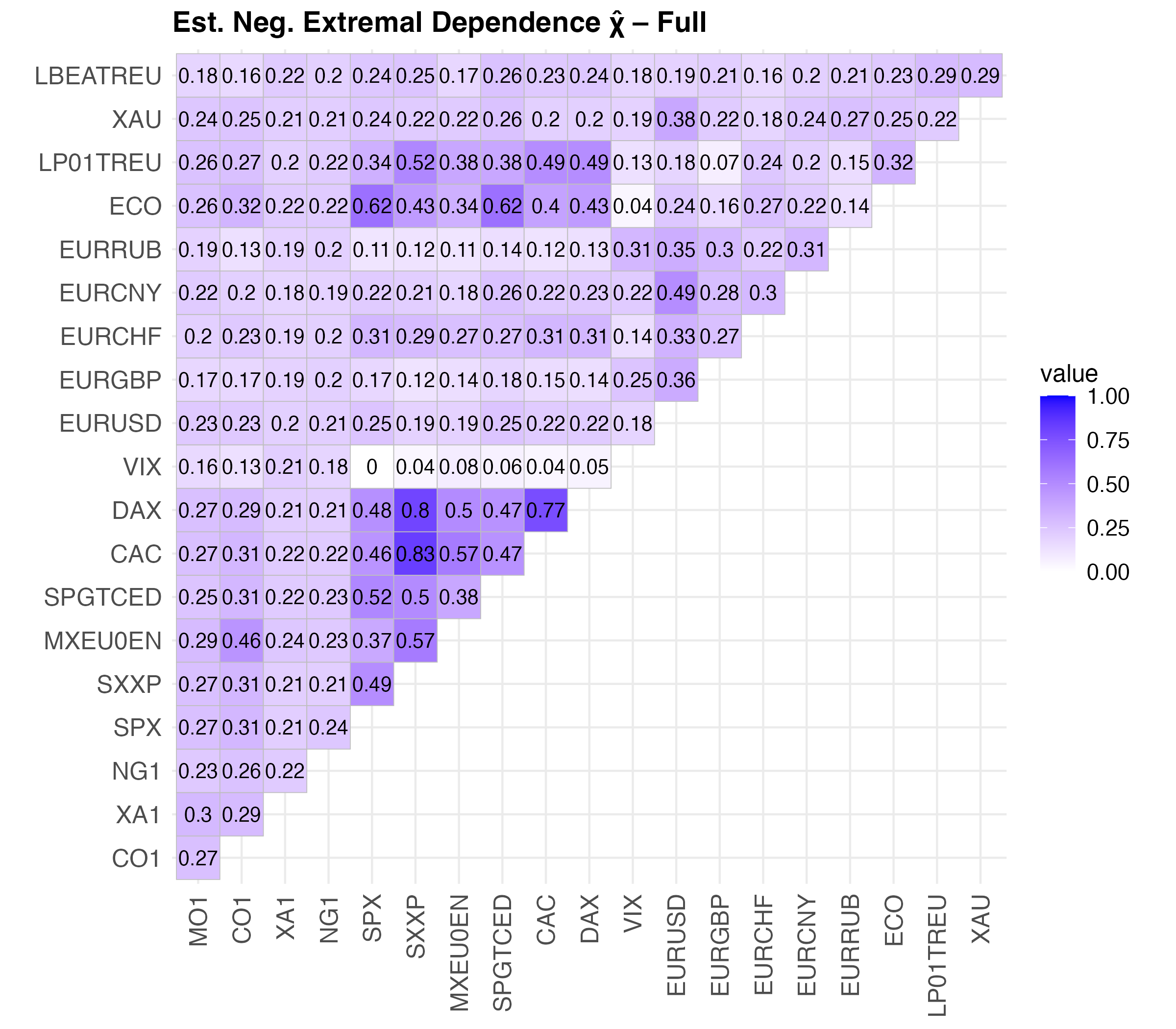}\\[6pt]
    \includegraphics[width=0.48\textwidth]{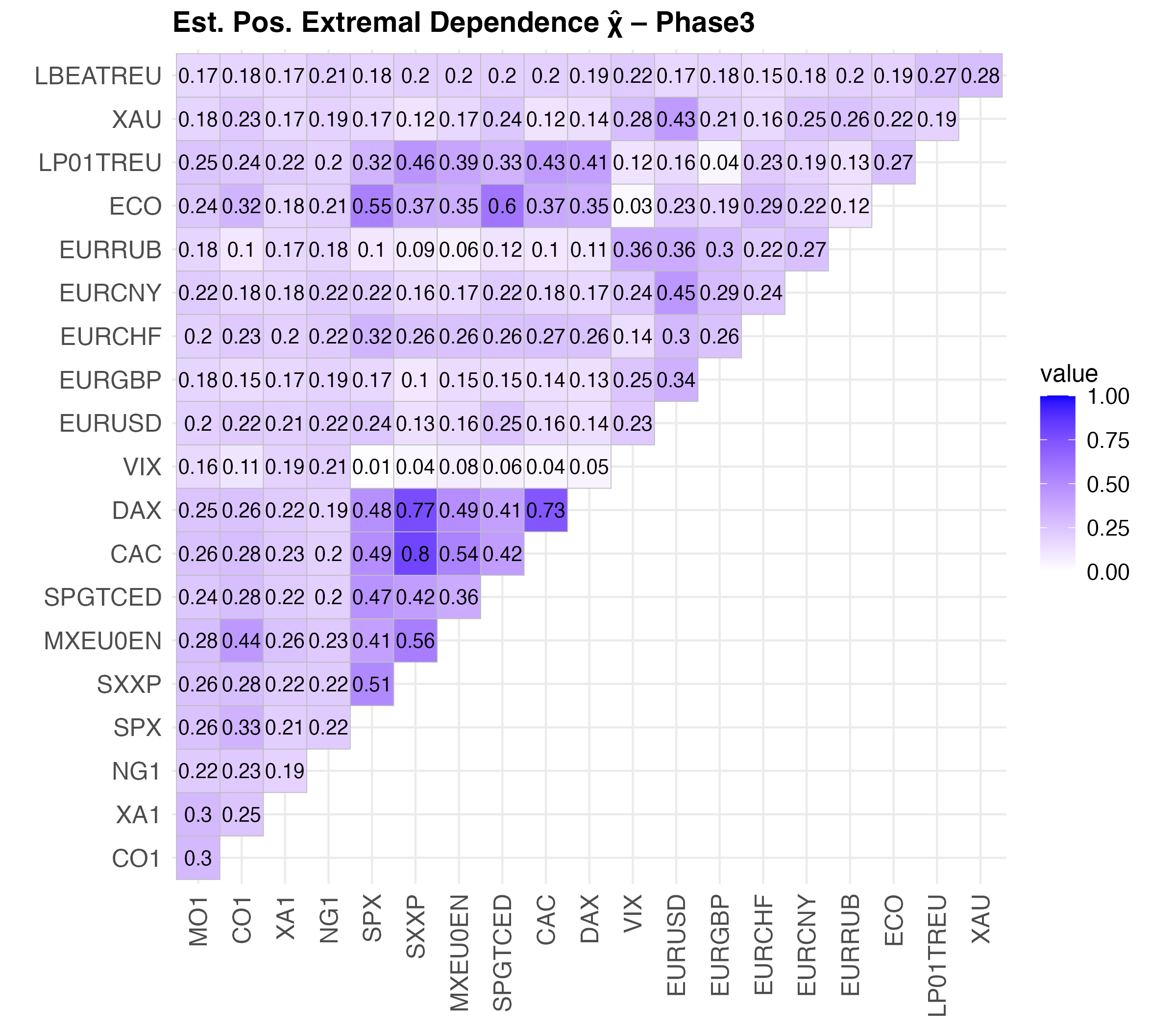}\hfill
    \includegraphics[width=0.48\textwidth]{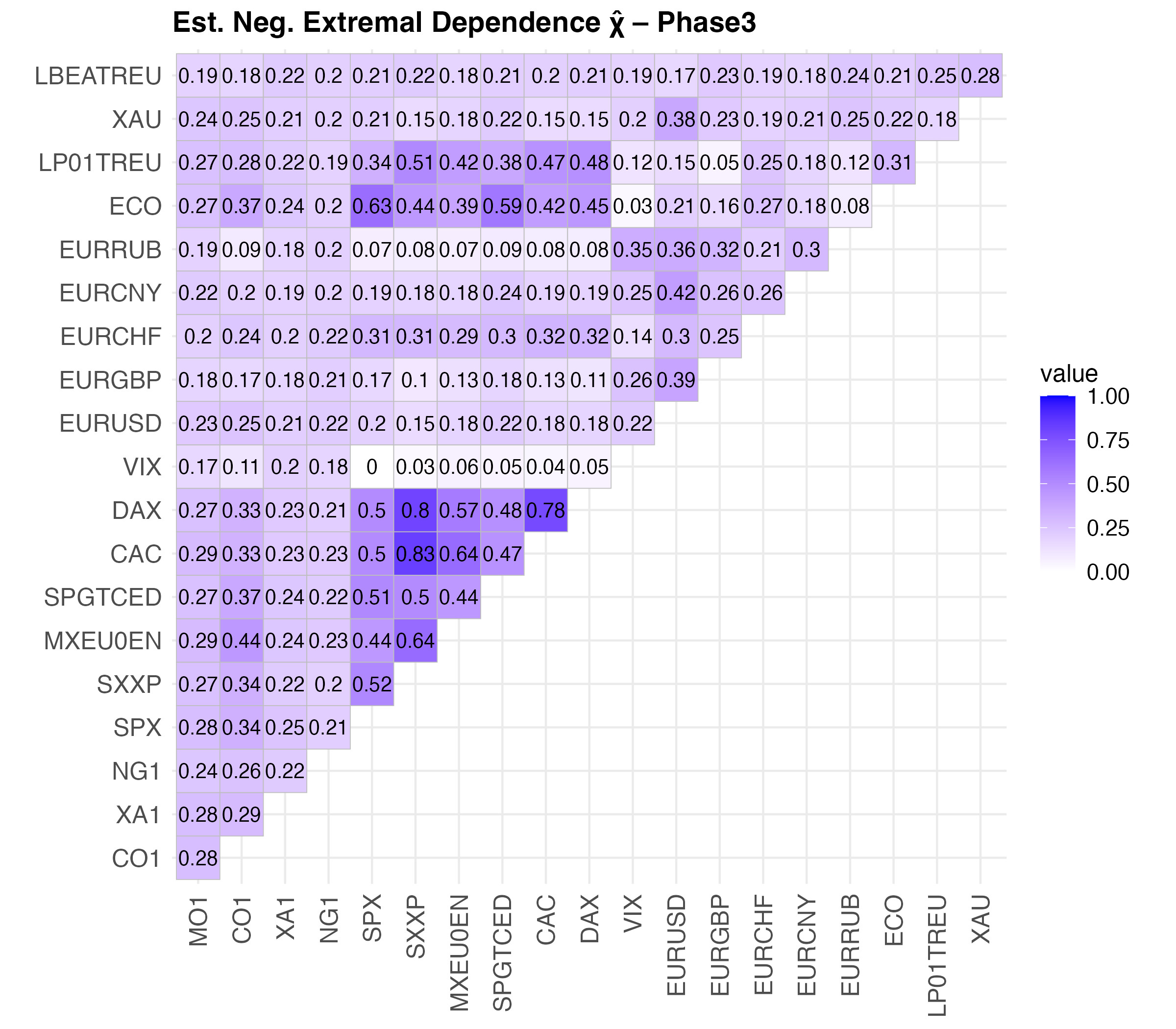}\\[6pt]
    \includegraphics[width=0.48\textwidth]{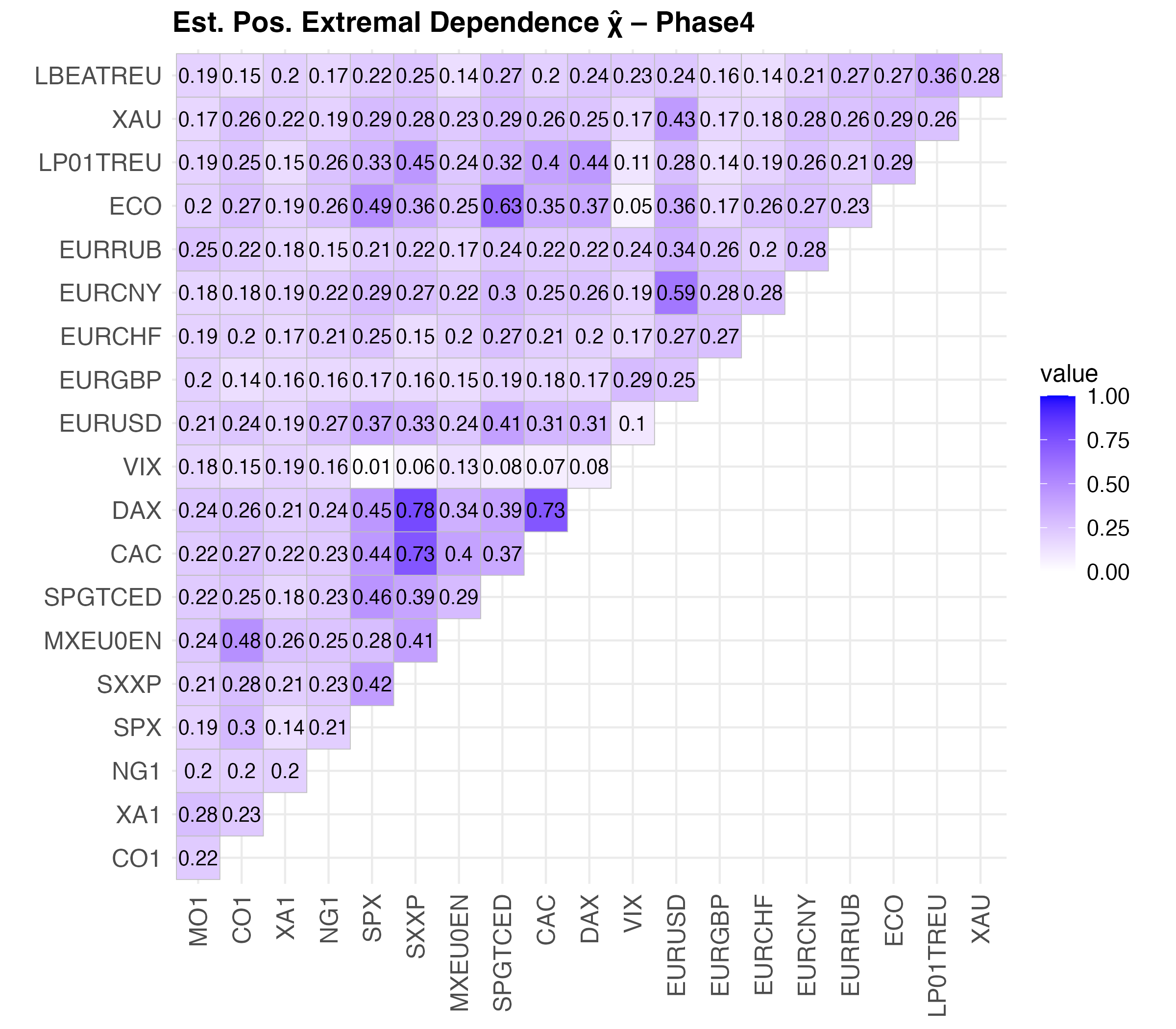}\hfill
    \includegraphics[width=0.48\textwidth]{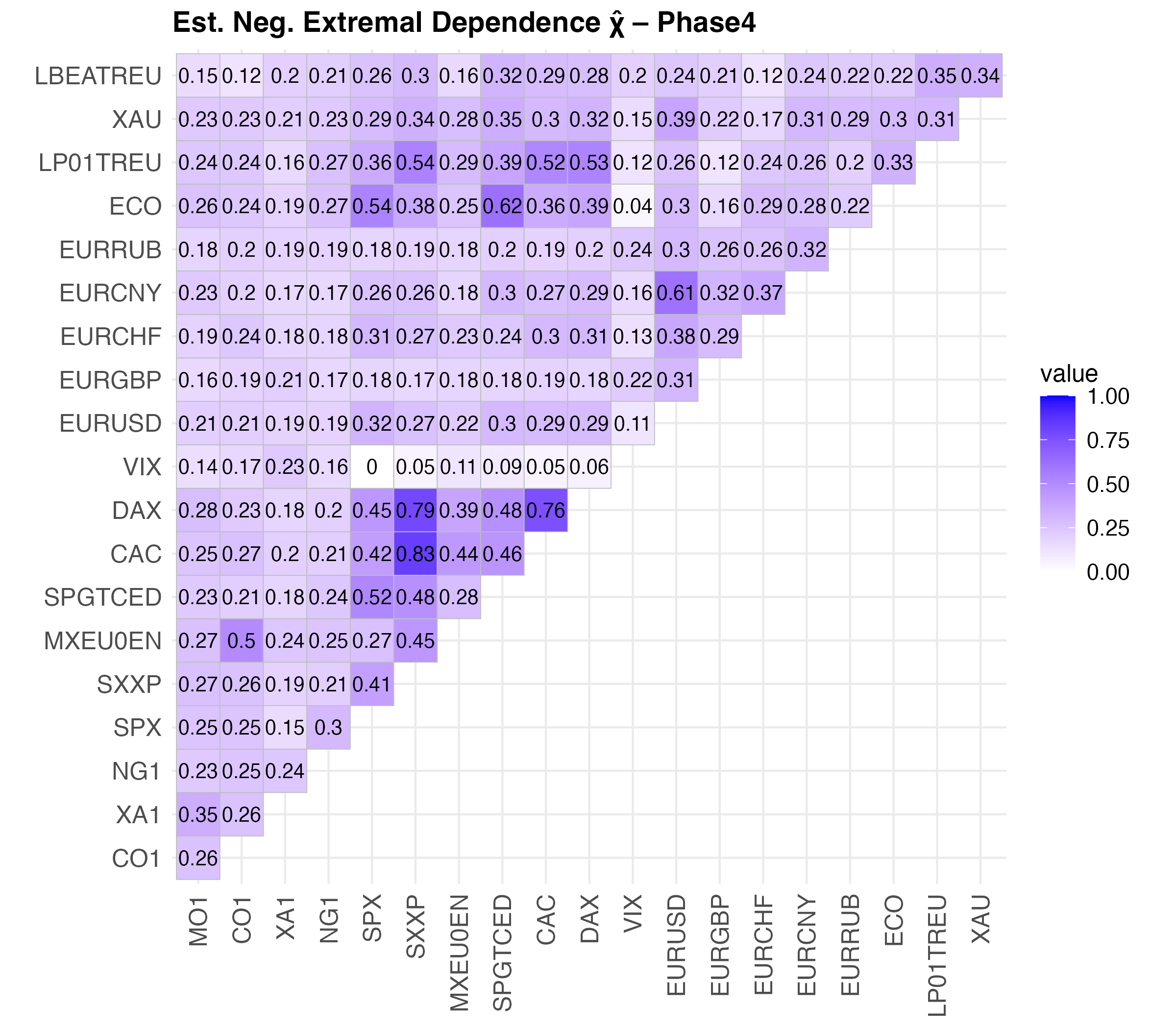}
    \caption{Estimated extremal dependence matrices $\hat{\chi}$. Left column:
    positive extremes. Right column: negative extremes. Rows: full sample
    (top), Phase~3 (middle), Phase~4 (bottom). Darker shading indicates
    stronger pairwise tail dependence ($\hat{\chi}$ approaching~1).}
    \label{fig:chi}
\end{figure*}

Figure~\ref{fig:chi} displays the estimated extremal dependence matrices
$\hat{\chi}$ for positive and negative extremes across all three sample
periods. These provide the unconditional pairwise tail dependence structure
before the graphical model imposes conditional independence constraints. The
MO1 row of each matrix is reported in full in Table~\ref{tab:app:chi} in the
Appendix.

Three features of the $\hat{\chi}$ matrices are noteworthy.

First, the strongest pairwise extremal dependence occurs consistently within
the equity cluster. The CAC--DAX pair reaches $\hat{\chi} \approx 0.73$--$0.78$
across both tail directions and all three sample periods, reflecting
near-lock-step behavior of the two major continental European equity indices
during extreme market conditions. Other within-equity pairs are also
prominent: DAX--SXXP and CAC--SXXP both at $\hat{\chi} \approx 0.78$--$0.83$,
MXEU0EN--SXXP at $\hat{\chi} \approx 0.45$--$0.64$, and ECO--SPGTCED at
$\hat{\chi} \approx 0.59$--$0.63$. This within-sector concentration of strong
unconditional tail dependence directly foreshadows the within-sector homophily
finding of the ERGM analysis in Section~\ref{sec:res:ergm}.

Second, EUA futures (MO1) exhibit uniformly weak pairwise extremal dependence
with all other variables. Across both tail directions and all sample periods,
$\hat{\chi}$ values for MO1 range from approximately 0.14 to 0.35, with no
partner variable exceeding 0.35 in any configuration. The strongest EUA tail
partners are coal (MO1--XA1, $\hat{\chi} \approx 0.28$--$0.35$) and oil
(MO1--CO1, $\hat{\chi} \approx 0.22$--$0.30$), with mild asymmetry across
tails: the MO1--XA1 link is strongest in the Phase~4 negative tail
($\hat{\chi} = 0.35$). The weak unconditional tail dependence between EUA and
its main energy counterparts contrasts sharply with MO1's near-complete
connectivity in the conditional extreme networks documented in
Section~\ref{sec:res:extreme}. This discrepancy is not contradictory: the
graphical model conditions on all other variables, so a conditional tail
dependence edge can arise even when unconditional pairwise tail dependence is
modest, provided the dependence between MO1 and a given partner is direct
rather than mediated through the rest of the system.

Third, MO1's $\hat{\chi}$ profile is remarkably stable across phases despite
the contraction of its conditional neighborhood in the Phase~4 negative
extreme network. While MO1's conditional tail neighborhood restructures between Phase~3 and Phase~4 negative extremes, its unconditional pairwise tail dependence values remain in the same 0.14--0.35 range throughout. The
Phase~3-to-Phase~4 restructuring of the tail network is therefore a change
in the conditional architecture of dependence (which variables mediate
extreme co-movement) rather than a change in the raw co-occurrence of
extreme returns.

%%-------------------------------------------------------------
\subsection{Divergence from average dependence}\label{sec:res:divergence}
%%-------------------------------------------------------------

\begin{figure*}
    \centering
    \includegraphics[width=\columnwidth]{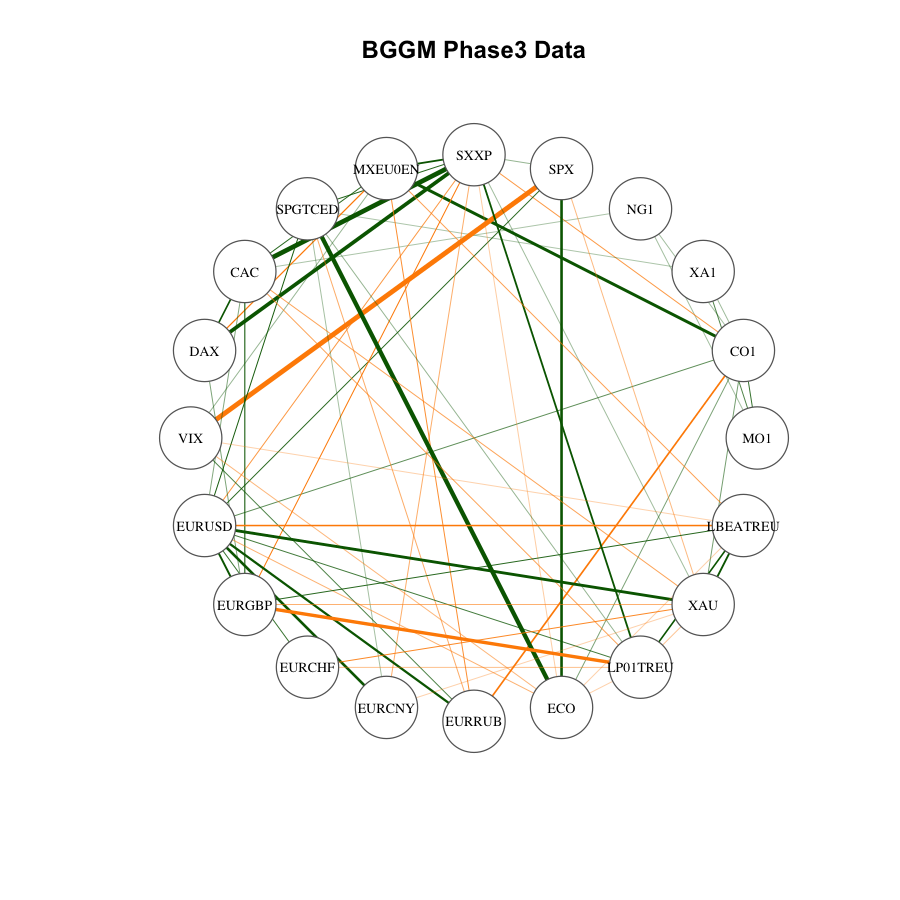}\hfill
    \includegraphics[width=\columnwidth]{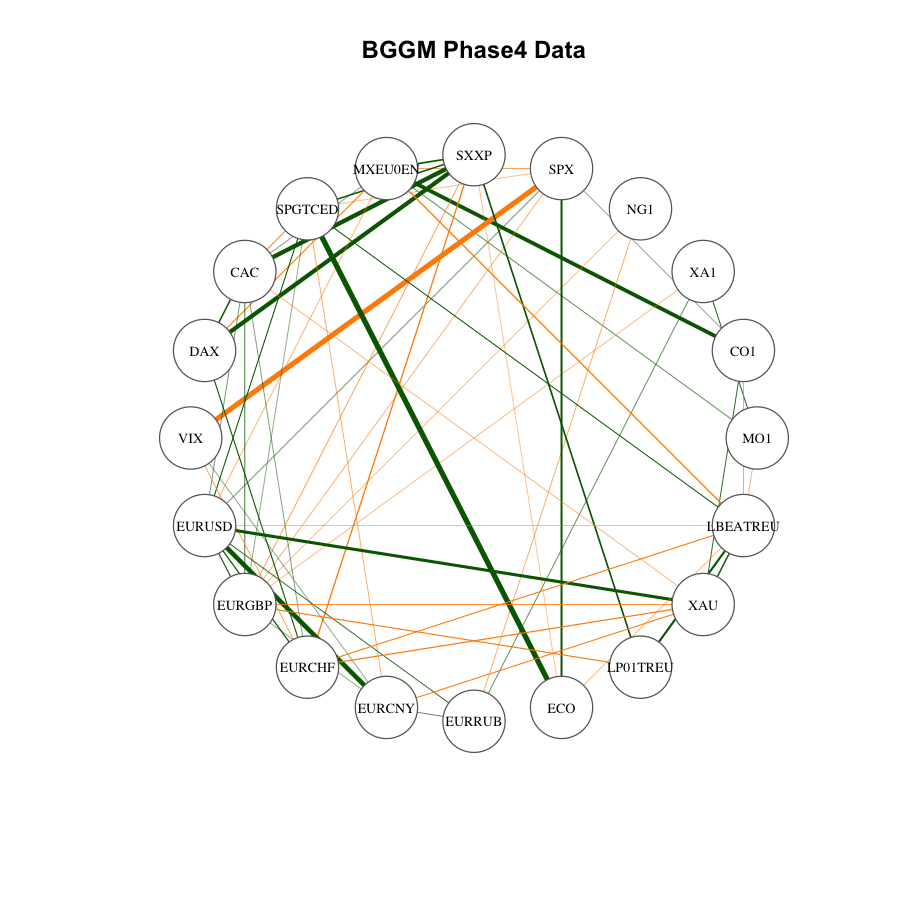}\\
    
    \vspace{-1.2cm}
    \includegraphics[width=\columnwidth]{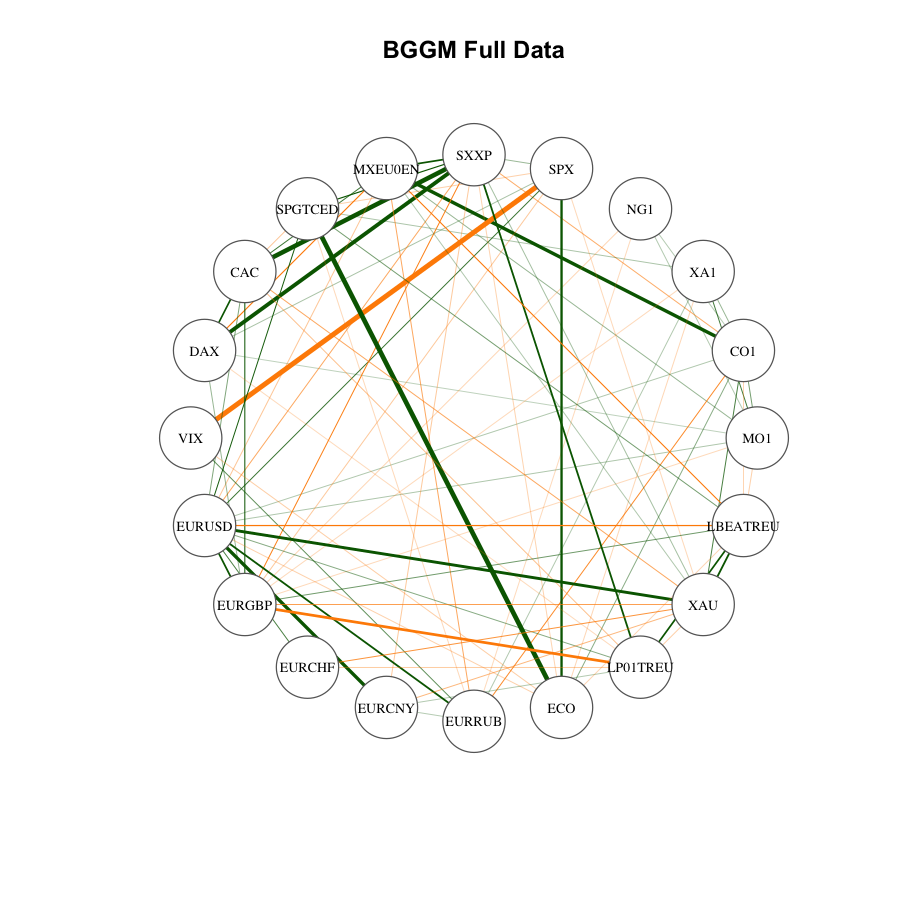}
    \vspace{-1.2cm}
    \caption{Standard Gaussian graphical models estimated via Bayesian
    inference (BGGM). Green edges indicate positive partial correlations,
    orange edges negative ones; edge width is proportional to partial
    correlation magnitude. Top/Left: Phase~3 (2013--2020). Top/Right: Phase~4 (2021--2025).
    Bottom: Full sample.}
    \label{fig:bggm}
\end{figure*}

We now contrast the extreme networks with the standard GGMs. Figure~\ref{fig:bggm} displays the
standard GGMs for the three sample periods, with green and orange edges
representing positive and negative partial correlations respectively.

The divergence between average and tail dependence operates along multiple
dimensions.

\paragraph{Density and connectivity.}
The extreme networks are substantially denser than the standard GGM across
all sample periods. In the full sample, the positive extreme network
(128 edges, density 0.674) contains 58\% more edges than the standard
network (81 edges, density 0.426). The gap is even larger in Phase~3 (127
versus 66 edges, a 92\% increase) and Phase~4 (119 versus 59 edges for
positive extremes, a 102\% increase). Although the relative decline in
density from Phase~3 to Phase~4 is similar across all three networks (10\%
for the standard GGM, 6\% for positive extremes, 11\% for negative
extremes), the standard network starts from a much lower base, so the
absolute gap between tail and average connectivity widens in Phase~4: the
ratio of positive extreme density to standard density rises from 1.93 in
Phase~3 to 2.01 in Phase~4. Risk models calibrated on average dependence
will therefore underestimate the connectivity of the tail network, and the
degree of underestimation grows in the more recent phase.

\paragraph{Centrality reversal.}
The most striking divergence is in the structural roles of individual nodes.
MO1 is among the least central nodes in the standard GGM: degree 3,
eigenvector centrality 0.12 in Phase~3, and 0.24 in Phase~4
(Table~\ref{tab:mo1}). Its conditional neighborhood in the standard GGM
contracts from eight variables in the full sample: CO1 ($+0.076$), XA1
($+0.103$), NG1 ($+0.041$), MXEU0EN ($+0.056$), DAX ($+0.035$), EURUSD
($+0.044$), EURGBP ($-0.037$), and LBEATREU ($-0.045$); to three in
Phase~3 (CO1, XA1, NG1) and three in Phase~4 (XA1, MXEU0EN, LBEATREU). In
the extreme networks, by contrast, MO1 connects to 13--18 of 19 possible
neighbors across all configurations and achieves eigenvector centrality of
0.91--1.00.

The reversal is symmetric for SXXP. In the standard GGM, SXXP is the
highest-degree node in Phase~3 (degree 12, eigenvector 0.95, betweenness
19.3) and remains highly central in Phase~4 (degree 8, eigenvector 0.74). In
every extreme network, SXXP has degree 6--7, the lowest in the system, with
eigenvector centrality 0.32--0.42. The STOXX Europe 600 thus transitions from
the most structurally important node under normal conditions to the most
peripheral node under stress. A similar pattern holds for EURUSD: it is the
highest-eigenvector node in the standard GGM across all periods (eigenvector
1.00 in all three) but drops to degree 8--9 and eigenvector 0.47--0.57 in
the positive extreme networks. The complete node-level centrality results for
all three network types are reported in
Tables~\ref{tab:app:centrality_std}--\ref{tab:app:centrality_neg} in the
Appendix.

\paragraph{Community structure.}
The standard GGM displays higher modularity (0.155--0.195) than the extreme
networks (0.070--0.119), with more communities (4--5 versus 2--3). The
standard network organizes cleanly along sector lines: in Phase~3, five
communities separate commodity, US equity, European equity, FX, and
bond-volatility groups. The extreme networks' lower modularity reflects the
diffuse nature of tail dependence: during extremes, the sector boundaries
visible in average dependence partially dissolve.

\paragraph{Phase transition asymmetry.}
The comparison between phases reveals a critical asymmetry. In the standard
GGM, the network thins from 81 edges in the full sample to 66 in Phase~3 and
59 in Phase~4, and MO1's neighborhood contracts accordingly. In the extreme
networks, the thinning is far more moderate: the Phase~4 positive extreme
retains 119 edges, twice the edge count of the Phase~4 standard GGM. The
widening gap between tail and average connectivity in Phase~4 has direct
implications for risk management: models calibrated on the average
conditional dependence structure of the current EU~ETS will substantially
underestimate the pervasiveness of tail co-movement.

%%-------------------------------------------------------------
%%-------------------------------------------------------------
\subsection{Structural drivers of tail connectivity}\label{sec:res:ergm}
%%-------------------------------------------------------------
To formalize the patterns identified above, we fit ERGMs to all learned
networks. We focus on the M2 specification, which includes an edges term,
sector-level nodefactor terms, a within-sector homophily term (nodematch),
and a triadic closure term (gwesp). M2 is preferred by AIC for all nine
networks and by BIC for six of the nine, with the simpler M0 preferred by
BIC for the remaining three. Table~\ref{tab:ergm_m2} reports M2 results
for all nine networks. Full
coefficient tables for the M0--M2 specifications across all nine networks,
together with AIC and BIC model selection statistics, are reported in
Tables~\ref{tab:app:ergm_m0}--\ref{tab:app:ergm_selection} in the Appendix.

\begin{table*}[htbp]
\centering
\caption{ERGM results for all nine networks under the M2 specification.
Bond is the reference sector for all nodefactor terms. Standard errors in
parentheses. Significance: $^{***}p<0.001$, $^{**}p<0.01$, $^{*}p<0.05$,
$^{.}p<0.1$.}\label{tab:ergm_m2}
\resizebox{\textwidth}{!}{
\begin{tabular}{l rrr rrr rrr}
\toprule
 & \multicolumn{3}{c}{Full} & \multicolumn{3}{c}{Phase~3} &
   \multicolumn{3}{c}{Phase~4} \\
\cmidrule(lr){2-4}\cmidrule(lr){5-7}\cmidrule(lr){8-10}
Term & Std & Pos& Neg & Std & Pos & Neg & Std & Pos & Neg \\
\midrule
edges
  & $-$1.53 (1.24)
  & $-$6.07 (5.01)
  & $-$59.74$^{*}$ (26.54)
  & $-$1.94$^{.}$ (1.03)
  & $-$15.26 (15.90)
  & $-$37.17$^{.}$ (21.28)
  & $-$0.70 (0.96)
  & $-$6.56 (7.48)
  & $-$0.91 (2.81) \\
nodefactor: Clean Eq.
  & $-$0.18 (0.46)
  & $-$0.64 (0.59)
  & 0.05 (0.46)
  & $-$0.11 (0.43)
  & $-$0.70 (0.55)
  & $-$0.59 (0.55)
  & 0.02 (0.54)
  & $-$0.10 (0.54)
  & 0.01 (0.52) \\
nodefactor: Commodity
  & $-$0.54 (0.40)
  & 0.03 (0.53)
  & 0.38 (0.45)
  & $-$0.81$^{*}$ (0.40)
  & 0.49 (0.48)
  & $-$0.04 (0.44)
  & $-$1.01$^{.}$ (0.52)
  & 0.05 (0.47)
  & 0.53 (0.47) \\
nodefactor: Equity
  & $-$0.25 (0.41)
  & $-$1.84$^{***}$ (0.54)
  & $-$1.88$^{***}$ (0.53)
  & $-$0.43 (0.38)
  & $-$2.04$^{***}$ (0.55)
  & $-$1.92$^{***}$ (0.52)
  & 0.02 (0.47)
  & $-$1.31$^{**}$ (0.50)
  & $-$0.59 (0.46) \\
nodefactor: FX
  & $-$0.34 (0.39)
  & $-$1.36$^{*}$ (0.54)
  & $-$0.28 (0.44)
  & $-$0.59 (0.39)
  & $-$1.08$^{*}$ (0.54)
  & $-$1.00$^{.}$ (0.52)
  & 0.14 (0.47)
  & $-$0.72 (0.49)
  & 0.25 (0.46) \\
nodefactor: Volatility
  & $-$0.39 (0.47)
  & $-$0.70 (0.58)
  & $-$0.08 (0.50)
  & $-$0.11 (0.41)
  & $-$0.74 (0.55)
  & $-$0.63 (0.55)
  & $-$0.28 (0.54)
  & $-$0.22 (0.54)
  & 0.13 (0.52) \\
nodematch
  & 1.31$^{**}$ (0.45)
  & 2.65$^{**}$ (0.86)
  & 3.69$^{***}$ (0.91)
  & 1.60$^{***}$ (0.46)
  & 3.94$^{***}$ (1.08)
  & 18.11 (1908)
  & 1.16$^{*}$ (0.50)
  & 2.86$^{***}$ (0.82)
  & 3.60$^{***}$ (1.08) \\
gwesp (0.5)
  & 0.77 (0.48)
  & 5.00$^{.}$ (2.93)
  & 36.35$^{*}$ (15.90)
  & 0.90$^{*}$ (0.38)
  & 10.32 (9.40)
  & 23.22$^{.}$ (12.63)
  & 0.004 (0.25)
  & 4.53 (4.33)
  & 0.47 (1.49) \\
\bottomrule
\end{tabular}}
\end{table*}

\paragraph{Within-sector homophily.}
The nodematch coefficient is positive and statistically significant in every
network without exception. In the extreme networks it ranges from 2.65
($p=0.002$) in the full positive extreme to 18.11 in Phase~3 negative
extremes. The standard GGM also exhibits significant homophily in every
period: 1.31 ($p=0.004$) in the full sample, 1.60 ($p<0.001$) in Phase~3,
and 1.16 ($p=0.020$) in Phase~4. Within-sector homophily is therefore a
universal structural feature of conditional dependence in this system,
whether average or extreme. However, the homophily coefficient is roughly
two to three times larger in extreme networks than in the standard GGM,
indicating that sector boundaries are substantially more binding in the
tails. This ERGM-based finding confirms the visual impression from the
$\hat{\chi}$ matrices, where the strongest pairwise tail dependence values
cluster within sectors.

\paragraph{Sector peripherality.}
The Equity sector nodefactor is the most consistently negative and
significant term across extreme networks. In the full sample it reaches
$-1.84$ ($p<0.001$) for positive and $-1.88$ ($p<0.001$) for negative
extremes; in Phase~3 it reaches $-2.04$ ($p<0.001$) and $-1.92$
($p<0.001$) respectively; in Phase~4 positive extremes it remains
significant at $-1.31$ ($p=0.009$). Equity nodes form systematically fewer
tail connections than expected given the overall density, even after
controlling for within-sector homophily. The FX sector shows a similar
peripheral pattern in the full sample and Phase~3 ($-1.36$, $p=0.011$ for
full positive extremes; $-1.08$, $p=0.046$ and $-1.00$, $p=0.055$ for
Phase~3 positive and negative extremes), weakening to non-significance in
Phase~4. By contrast, no sector nodefactor reaches significance in any
period of the standard GGM, except for the Phase~4 Commodity term ($-1.01$,
$p=0.052$), indicating that sectors are broadly equally connected at the
average-dependence level. These results formalize the visual observation
that equity indices (particularly SXXP) and FX nodes sit at the edges of
the extreme networks.

A striking contrast emerges in Phase~4 negative extremes, where no
nodefactor term is significant. The Phase~4 crash-tail network has become
sectorally homogeneous: no sector is systematically more or less connected
than expected. The Equity peripherality that characterizes every other
extreme network dissolves in the Phase~4 negative tail, suggesting that the
energy crisis and post-COVID market regime have equalized crash-risk
exposure across sectors.

The Phase~4 standard GGM presents a mirror-image pattern. The Commodity
nodefactor is the only marginally significant sector term ($-1.01$,
$p=0.052$), while Equity is essentially zero ($0.02$, ns). In average
dependence, it is commodity nodes (not equity) that are structurally
peripheral in Phase~4. The structural role of sectors is thus inverted
between average and tail dependence: sectors central under normal conditions
are peripheral under stress, and vice versa.

\paragraph{Triadic closure.}
The gwesp term captures triadic closure: the tendency for shared neighbors
to be connected, creating triangular subgraphs through which shocks can
propagate in a clustered pattern. Triadic closure is strongest in the
negative tails. In the full-data negative extreme network, gwesp is $36.35$
($p=0.022$), indicating significant clustered contagion during simultaneous
market crashes. Phase~3 negative extremes show a similar but marginally
significant pattern ($23.22$, $p=0.066$). In positive extremes the picture
is more mixed: the full positive extreme estimate is marginally
significant ($5.00$, $p=0.088$), while the Phase~3 and Phase~4 positive
extreme gwesp values (10.32 and 4.53) are non-significant. Among standard
networks, the Phase~3 standard GGM is the one average-dependence network to
exhibit significant triadic closure ($0.90$, $p=0.018$), with the much
smaller magnitude consistent with the milder clustering implied by
average-dependence covariation.

Phase~4 eliminates triadic closure entirely. In the Phase~4 negative
extreme network, gwesp is $0.47$ (ns), in the Phase~4 positive extreme
$4.53$ (ns), and in the Phase~4 standard GGM $0.004$ (ns). This is
consistent with the nature of Phase~4 market stress: the persistent energy
supply disruptions and gradual geopolitical risk build-up of 2021--2023
propagated through a more diffuse tail structure, in contrast with the
sharp synchronous dislocation of the March 2020 COVID-19 crash, a canonical
clustered contagion event, which falls squarely within the Phase~3 window.

%%=============================================================
\section{Discussion}\label{sec:discussion}
%%=============================================================

The results presented in Section~\ref{sec:results} establish a series of
findings about the conditional tail dependence structure of EUA futures
markets that, taken together, recast several conclusions from the existing
literature. We discuss four themes in turn: the architecture of risk under
stress, the Phase~3-to-Phase~4 transition, clustered versus diffuse
contagion, and the implications for risk management and regulation.

\subsection{The architecture of risk under stress}

The most fundamental finding is that the network of tail dependencies in
the EU carbon market is structurally distinct from the network of average
dependencies. This is consistent with, and extends, the quantile
connectedness results of \citet{wei2023} and \citet{cao2024}, who document
that marginal connectedness measures among EUA and energy markets rise from
approximately 28\% at the median to 77\% at the tails. Our conditional
graphical model framework recovers an analogous, but more precise,
pattern: extreme network densities of 0.58--0.67 against standard GGM
densities of 0.31--0.43, with a density ratio between 1.93 and 2.01
depending on phase. Where \citet{wei2023} and \citet{cao2024} measure
the volume of marginal spillover, we identify which specific conditional
edges are present in the tails and absent in the centre, and vice versa.

The reversal of node centrality between average and tail networks is, to our
knowledge, novel in the EUA literature. The STOXX Europe 600 (SXXP) and the
EURUSD exchange rate are the most central nodes in the standard GGM, with
eigenvector centrality of 0.95 and 1.00 in Phase~3 respectively; both drop
to peripheral positions in every extreme network configuration. EUA futures
themselves are peripheral in the standard GGM (degree 3 in both Phase~3 and
Phase~4) yet achieve the highest eigenvector centrality in extreme networks.
This pattern qualifies the implication, drawn from the structural equation
model of \citet{wang2020} and the Information Imbalance analyses of
\citet{salvagnin2024,salvagnin2025}, that financial and equity variables are
the dominant drivers of EUA pricing in Phase~4. Those analyses recover
average dependence structures, in which equity centrality is genuine. In the
tails, however, the relationship reverses: financial nodes are structurally
absent, and the dominant edges run between EUA, its commodity counterparts,
and the bond/volatility complex.

This reversal is consistent with the asymmetric lower tail dependence
between EUA and the WilderHill Clean Energy Index (ECO) documented by
\citet{hanif2021a}, the only previously identified statistically significant
tail dependence result involving EUA in the bivariate literature. Our
conditional analysis extends this in two directions. First, the EUA--ECO
edge is present in the full-sample and Phase~3 negative extreme networks but
disconnects in Phase~4 negative extremes, suggesting that the clean energy
tail link documented by \citet{hanif2021a} on 2011--2020 data weakens in the
most recent regulatory period. Second, the conditional analysis recovers a
much wider tail neighbourhood for EUA than any bivariate study could
identify: 17--18 edges in the full sample, spanning commodity, equity,
volatility, FX, and bond variables.

The contrast with \citet{reboredo2015}, who found Gaussian copulas best fit
EUA--oil and EUA--gas pairs in Phase~2 (implying no tail dependence with
fossil fuels), is striking. Our Phase~4 negative extreme network identifies
MO1--XA1 (coal) and MO1--CO1 (oil) as the strongest EUA tail partners, with
$\hat{\chi}$ reaching 0.35 for MO1--XA1. The discrepancy is most plausibly
explained by the deepening financialization of EUA markets after Phase~2:
during the early years of the EU ETS, EUA prices were sufficiently
disconnected from energy markets that even bivariate tail dependence was
absent, whereas by Phase~3--4 the markets are sufficiently integrated for
strong tail co-movement to emerge. This also answers, in the multivariate
direction explicitly suggested by \citet{feng2012}, their call to extend
the univariate EVT framework for EUA risk assessment.

\subsection{The Phase 3-to-Phase 4 transition}

The phase comparison reveals what is, to our reading, the most policy-relevant
finding: average dependence contracts sharply in Phase~4 while tail
dependence persists. The standard GGM thins from 81 edges in the full sample
to 66 in Phase~3 and 59 in Phase~4, a 27\% decline in density. The extreme
networks contract far less, with positive extreme density falling only 6\%
and negative extreme density 11\%. The ratio of extreme to standard density
therefore rises from 1.93 in Phase~3 to 2.01 in Phase~4.

This finding synthesizes and qualifies several existing results.
\citet{dittmann2025} document that energy fundamentals explain 12\% of EUA
price variance in Phase~3 but below 1\% in Phase~4 via variance decomposition,
and \citet{salvagnin2024} find that Phase~3 EUA prices are dominated by
energy variables while Phase~4 is dominated by financial and currency
variables via the Information Imbalance criterion. Both findings describe
the average dependence structure. Our results show that this apparent
decoupling of EUA from energy fundamentals in Phase~4 is a feature of the
average dependence channel only: in the tails, the connections between EUA
and coal, oil, and gas persist with similar strength to Phase~3. EUA's tail
neighborhood in Phase~4 contains CO1, XA1, and NG1 in both directions of the
tail, exactly as in Phase~3. The financialization narrative documented by
\citet{borghesi2023} and \citet{terranova2025} is therefore incomplete: while
the average dependence structure has indeed shifted toward financial and
sentiment-driven dynamics, the tail dependence channel remains commodity-anchored.

The asymmetric restructuring of EUA's tail neighborhood between Phase~3 and
Phase~4 is consistent with \citet{berrisch2023}, who document the EUA--gas
correlation flipping from $+0.3$ to $-0.4$ around the Russian invasion of
Ukraine. In our negative extreme networks, the FX neighborhood of EUA
expands in Phase~4 with EURCHF and EURGBP joining, while equity and clean
energy nodes (DAX, ECO, SPGTCED) disconnect. The Phase~4 negative extreme
network is the only configuration with positive degree assortativity
($+0.039$), and EURRUB attains the highest eigenvector centrality (1.000),
reflecting the prominence of Russia-related FX in the post-invasion tail
structure. This network-level evidence complements the time-varying copula
finding of \citet{berrisch2023}: the structural break visible in their
EUA--gas correlation is part of a broader reorganization of the entire tail
network around the energy crisis.

The Equity sector peripherality identified by the ERGM analysis is robust
across all extreme networks except Phase~4 negative, where it dissolves
together with all other sector effects. Phase~4 negative extremes form a
sectorally homogeneous network: no sector is systematically more or less
connected than expected given overall density. This pattern is unique among
the nine networks we estimate and suggests that the persistent energy
supply disruptions and gradual geopolitical risk build-up of Phase~4 have
equalized crash-risk exposure across all sectors, in a way that neither
average dependence nor positive tail dependence shows.

\subsection{Clustered versus diffuse contagion}

The triadic closure finding represents a structural distinction between
two regimes of crash contagion that has not, to our knowledge, been
identified in the EUA literature. In the full-sample negative extreme
network, the gwesp coefficient is $36.35$ ($p = 0.022$): connected nodes
share many common neighbors, producing the triangular subgraphs through
which clustered contagion propagates. In Phase~3 negative extremes the
pattern is similar though marginally significant ($23.22$, $p = 0.066$).
In Phase~4 negative extremes the effect vanishes ($0.47$, ns), as it does in
the Phase~4 positive extreme and the Phase~4 standard networks.

This distinction maps cleanly onto the financialization literature.
\citet{borghesi2023} and \citet{terranova2025} emphasize that Phase~4
markets are increasingly driven by regulatory news, speculative activity,
and gradual sentiment shifts rather than by synchronous fundamentals shocks.
\citet{terranova2025} identifies seven speculative bubbles in EUA futures
between 2017 and 2022, six of which were triggered by regulatory
announcements rather than commodity fundamentals. Clustered contagion (shocks propagating through tightly connected neighborhoods to all nearby
nodes simultaneously) is the natural network signature of a sharp,
synchronous fundamentals shock such as the March 2020 COVID-19 crash, which
falls squarely within the Phase~3 window. Diffuse propagation (shocks
moving through extended pathways without triangular clustering) is the
natural signature of a gradual, news- or sentiment-driven process such as
the 2021--2023 energy supply build-up and the Russo-Ukrainian conflict
escalation. The disappearance of triadic closure in Phase~4 is therefore
not an artefact of the smaller Phase~4 sample, but a structural feature
that aligns with the qualitative change in Phase~4 market dynamics
documented by the financialization literature.

Three features of Phase 4 market microstructure plausibly mediate this shift. First, the rise of algorithmic and high-frequency participation (the 73\% financial-intermediary share documented by ESMA for 2023) tends to spread shocks across larger numbers of correlated positions rather than concentrating them in identifiable counterparty triangles. Second, the MSR mechanism, by adjusting supply gradually rather than discretely, smooths the shock-propagation path: instead of a sharp synchronous adjustment that would generate clustered contagion, supply tightening unfolds over weeks. Third, the rolling sequence of regulatory announcements over 2021--2023, identified by \citet{terranova2025} as the dominant trigger of Phase 4 bubbles, produces a stream of partial shocks that diffuse through extended pathways rather than a single synchronous dislocation. 

\subsection{Implications for risk management and regulation}

Three practical implications follow from the findings above.

First, the divergence between average and tail dependence structures means
that standard graphical models are systematically misleading for tail risk
assessment. Risk models calibrated on average dependence (whether
correlation-based, partial-correlation-based, or based on the standard
graphical lasso) will identify equity indices and major FX pairs as the
dominant connections to EUA pricing, when in fact these are precisely the
nodes that disconnect under stress. For compliance entities and
institutional EUA holders constructing tail-event hedges, our results imply
that hedge selection should prioritize within-sector commodity exposure
(specifically coal and natural gas, with which EUA shares the strongest
conditional tail dependence) over cross-sector positions in equity indices
or currency pairs. This is consistent with the practitioner intuition that
fuel-switching commodities are the natural EUA hedges, but it is now
supported by a formal conditional dependence analysis at the tail.

Second, the widening density gap between extreme and average networks in
Phase~4 has direct implications for stress-testing frameworks. Any framework that calibrates interdependence parameters on average correlations or partial correlations, rather than on tail-quantile dependence, 
will increasingly underestimate the systemic exposure of ETS-regulated
institutions in the current regulatory environment. The fact that the
average-dependence channel has weakened while the tail-dependence channel
has persisted means that the most dangerous co-movements occur precisely
when stress tests need to capture them most accurately. Calibrating
stress-test interdependence parameters on extreme quantile data, or
explicitly on graphical models of extremes, would mitigate this risk.

Third, the disappearance of triadic closure in Phase~4 has implications
for systemic risk monitoring. Clustered contagion is easier to detect and
contain than diffuse propagation: a triangular subgraph of connected nodes
implies a small set of pivotal players whose stabilization can interrupt
the contagion path. Diffuse propagation through extended, weakly clustered
pathways is harder to interrupt because there are no natural intervention
points. The shift from clustered to diffuse tail dynamics in Phase~4
therefore suggests that traditional macroprudential tools designed for the
2020 crisis pattern may be less effective against the slow-burn stress
patterns of the 2021--2023 energy crisis, and that the next iteration of
ETS regulatory design (the planned MSR adjustments under the Fit for 55
package, and the integration of road transport and buildings into ETS2 scheduled for 2027 with a possible deferral to 2028) should be evaluated against tail-dependence structure rather
than against average-dependence structure alone.

\subsection{Limitations and avenues for future research}

Two broad classes of limitation should be acknowledged.

The first concerns the data and the variable set. Our analysis considers
twenty daily series spanning commodity, equity, clean equity, volatility,
FX, and bond categories, selected on the basis of the prior EUA literature
and our companion work \citep{maciejowski2026}. Other variables plausibly
relevant to EUA pricing are not included: electricity prices, weather and
temperature variables, broader macroeconomic indicators (industrial
production, inflation expectations), firm-level compliance positions, and
explicit measures of the regulatory environment such as MSR auction
withholding volumes, Fit-for-55 announcement dummies, or text-based
policy-uncertainty indices constructed from ETS-related communications.
The latter category is particularly relevant given that
\citet{terranova2025} attributes six of the seven speculative bubbles
detected in EUA futures over 2017--2022 to regulatory announcements rather
than commodity fundamentals; an explicit regulatory channel could plausibly
mediate some of the tail dependencies we attribute to other variables. The
omission of any genuinely important driver could in principle reorganize
the conditional dependence structure recovered here, since edges in a
graphical model are estimated conditional on the rest of the variable set,
and the strength of our findings should therefore be understood as
conditional on the variables included. Daily frequency also constrains
what the analysis can detect: intraday propagation of news shocks within
the trading day is invisible at this temporal resolution, as are the
high-frequency speculative dynamics emphasized by \citet{lovcha2021}, who
find that high-frequency EUA variation is dominated by speculative activity
while business-cycle frequencies are dominated by fundamentals. Tail
dependence at hourly or sub-hourly resolution may differ substantively from
the daily structure we recover. A related concern is that our Phase 4 sample jointly identifies the post-2020 regulatory regime, the COVID-19 recovery, the 2021--2022 energy crisis, and the Russo-Ukrainian war. The structural changes documented here should therefore be read as features of the Phase 4 environment as a whole, not as causal effects of the regulatory transition alone.

The second class of limitation is methodological, and three natural
extensions follow from the novelty of this application. First, the analysis is descriptive rather than
predictive; embedding the extreme graphical model in a dynamic framework
would enable probabilistic forecasting of the tail network structure,
with direct relevance to real-time stress-testing and macroprudential
supervision. Second, we analyzed Phases 3 and 4 independently, which
discards potential information sharing across regulatory periods; a
Bayesian hierarchical extension in which phase-specific parameters share a
common prior would borrow strength across phases while still allowing
genuine phase-specific differences to emerge, and would be especially
valuable for Phase~4 whose shorter sample limits precision. Third, the
phase split at December 2020 was imposed exogenously on the basis of the
regulatory calendar; changepoint detection methods for extremes have been
developed in the univariate and bivariate settings
\citep[e.g.][]{de2020tracking,lattanzi2021change}, and extending them to the multivariate
graphical-model framework would allow structural breaks in tail dependence
to be inferred from the data rather than imposed.
%%=============================================================
%%=============================================================
\section{Conclusion}\label{sec:conclusion}
%%=============================================================

This paper has applied graphical models of extremes to EUA futures data
spanning Phases 3 and 4 of the EU ETS, characterizing the conditional tail dependence structure of the EU carbon market.
Across the full sample and each phase separately, extreme networks are
structurally distinct from the standard Gaussian graphical model: they are
substantially denser, organized around different central nodes, and
governed by within-sector homophily that binds sector boundaries more
tightly than at the average-dependence level. EUA futures themselves are
peripheral in the standard GGM but achieve the highest centrality in
extreme networks, while equity indices and major FX pairs follow the
opposite trajectory. The Phase~3-to-Phase~4 transition reveals an
asymmetric restructuring: average dependence contracts sharply in the
post-2020 regulatory period while tail dependence persists, and the
clustered triadic contagion characterizing Phase~3 negative extremes gives
way to a more diffuse propagation structure in Phase~4 consistent with the
gradual, news-driven stress patterns of the energy crisis.

Together, these findings support a view of the EU carbon market in which
the architecture of risk under stress differs fundamentally from the
architecture of co-movement in normal conditions. Standard graphical
models, however carefully estimated, provide a systematically misleading
guide for tail risk assessment, hedge construction, and regulatory
stress-testing. The extreme graphical model framework introduced here
offers a more appropriate tool for these purposes, and the ERGM analysis
of learned network topology provides a principled way to compare
dependence structures across market regimes. As the EU ETS evolves through
further regulatory tightening and the planned integration of additional
sectors, tail dependence rather than average dependence should be the
primary lens through which systemic risk is monitored and managed.

\printcredits

\bibliographystyle{cas-model2-names}

\bibliography{cas-refs}

%%=============================================================
\appendix
%%=============================================================

\section{Supplementary network statistics}\label{app:stats}

This appendix reports the complete numerical results underlying the main
text: node-level centrality measures for all three network types and all
three sample periods, the full row of MO1 in the $\hat{\chi}$ matrix across
all six configurations, and the complete sequence of ERGM specifications
together with model selection statistics.

\begin{table*}[htbp]
\centering
\caption{Node-level centrality measures for the standard GGM across sample
periods. Deg = degree; Betw = betweenness; Eigen = eigenvector centrality;
PR = PageRank.}\label{tab:app:centrality_std}
\small
\begin{tabular}{lrrrrrrrrrrrr}
\toprule
 & \multicolumn{4}{c}{Full} & \multicolumn{4}{c}{Phase~3} &
   \multicolumn{4}{c}{Phase~4} \\
\cmidrule(lr){2-5}\cmidrule(lr){6-9}\cmidrule(lr){10-13}
Node & Deg & Betw & Eigen & PR & Deg & Betw & Eigen & PR & Deg & Betw & Eigen & PR \\
\midrule
MO1      & 8  & 4.1  & 0.547 & 0.050 & 3  & 0.7  & 0.119 & 0.031 & 3  & 3.3  & 0.243 & 0.030 \\
CO1      & 10 & 7.6  & 0.683 & 0.060 & 9  & 30.6 & 0.602 & 0.070 & 4  & 0.8  & 0.413 & 0.035 \\
XA1      & 6  & 2.2  & 0.390 & 0.039 & 3  & 1.8  & 0.161 & 0.029 & 3  & 3.9  & 0.199 & 0.032 \\
NG1      & 4  & 0.8  & 0.269 & 0.028 & 3  & 2.2  & 0.175 & 0.029 & 2  & 0.6  & 0.164 & 0.023 \\
SPX      & 8  & 7.8  & 0.571 & 0.050 & 5  & 1.8  & 0.477 & 0.038 & 7  & 10.9 & 0.651 & 0.058 \\
SXXP     & 12 & 8.6  & 0.858 & 0.070 & 12 & 19.3 & 0.947 & 0.084 & 8  & 10.6 & 0.742 & 0.065 \\
MXEU0EN  & 9  & 4.5  & 0.678 & 0.054 & 7  & 9.1  & 0.489 & 0.053 & 8  & 12.9 & 0.730 & 0.065 \\
SPGTCED  & 8  & 4.1  & 0.565 & 0.049 & 7  & 9.6  & 0.552 & 0.053 & 8  & 6.9  & 0.842 & 0.063 \\
CAC      & 8  & 1.8  & 0.639 & 0.048 & 8  & 11.0 & 0.661 & 0.059 & 8  & 6.2  & 0.863 & 0.063 \\
DAX      & 7  & 2.2  & 0.517 & 0.043 & 4  & 0.3  & 0.348 & 0.032 & 4  & 0.2  & 0.431 & 0.035 \\
VIX      & 3  & 0.5  & 0.172 & 0.024 & 5  & 2.1  & 0.331 & 0.039 & 3  & 1.2  & 0.273 & 0.029 \\
EURUSD   & 15 & 22.5 & 1.000 & 0.087 & 13 & 23.3 & 1.000 & 0.091 & 10 & 17.4 & 1.000 & 0.078 \\
EURGBP   & 11 & 12.5 & 0.760 & 0.065 & 7  & 2.3  & 0.654 & 0.051 & 10 & 32.8 & 0.872 & 0.082 \\
EURCHF   & 3  & 0.0  & 0.268 & 0.023 & 3  & 0.0  & 0.330 & 0.026 & 7  & 8.8  & 0.688 & 0.057 \\
EURCNY   & 5  & 1.1  & 0.413 & 0.033 & 4  & 0.3  & 0.420 & 0.032 & 6  & 6.7  & 0.571 & 0.051 \\
EURRUB   & 8  & 11.6 & 0.458 & 0.052 & 5  & 3.5  & 0.376 & 0.040 & 4  & 4.6  & 0.276 & 0.040 \\
ECO      & 8  & 6.1  & 0.541 & 0.050 & 8  & 7.3  & 0.671 & 0.058 & 4  & 0.7  & 0.429 & 0.035 \\
LP01TREU & 9  & 5.1  & 0.651 & 0.055 & 9  & 6.6  & 0.789 & 0.064 & 3  & 0.7  & 0.341 & 0.028 \\
XAU      & 11 & 8.1  & 0.791 & 0.065 & 10 & 12.6 & 0.820 & 0.071 & 7  & 6.5  & 0.739 & 0.056 \\
LBEATREU & 9  & 3.9  & 0.683 & 0.054 & 7  & 4.8  & 0.602 & 0.051 & 9  & 17.4 & 0.774 & 0.073 \\
\bottomrule
\end{tabular}
\end{table*}

\begin{table*}[htbp]
\centering
\caption{Node-level centrality measures for the positive extreme GGM across
sample periods.}\label{tab:app:centrality_pos}
\small
\begin{tabular}{lrrrrrrrrrrrr}
\toprule
 & \multicolumn{4}{c}{Full} & \multicolumn{4}{c}{Phase~3} &
   \multicolumn{4}{c}{Phase~4} \\
\cmidrule(lr){2-5}\cmidrule(lr){6-9}\cmidrule(lr){10-13}
Node & Deg & Betw & Eigen & PR & Deg & Betw & Eigen & PR & Deg & Betw & Eigen & PR \\
\midrule
MO1      & 17 & 4.9  & 1.000 & 0.064 & 16 & 4.5  & 0.961 & 0.061 & 16 & 6.9  & 1.000 & 0.065 \\
CO1      & 11 & 0.8  & 0.701 & 0.044 & 14 & 2.4  & 0.868 & 0.054 & 11 & 1.2  & 0.758 & 0.046 \\
XA1      & 16 & 3.9  & 0.954 & 0.060 & 17 & 6.3  & 0.996 & 0.064 & 12 & 2.5  & 0.785 & 0.050 \\
NG1      & 16 & 3.7  & 0.959 & 0.060 & 17 & 5.8  & 1.000 & 0.064 & 15 & 4.9  & 0.965 & 0.061 \\
SPX      & 16 & 7.0  & 0.915 & 0.061 & 17 & 9.4  & 0.959 & 0.065 & 13 & 5.6  & 0.794 & 0.054 \\
SXXP     & 6  & 0.3  & 0.339 & 0.028 & 6  & 0.1  & 0.322 & 0.028 & 7  & 1.1  & 0.424 & 0.033 \\
MXEU0EN  & 11 & 2.7  & 0.654 & 0.044 & 11 & 2.4  & 0.656 & 0.045 & 10 & 3.1  & 0.644 & 0.043 \\
SPGTCED  & 13 & 4.3  & 0.754 & 0.051 & 13 & 5.7  & 0.730 & 0.052 & 11 & 3.1  & 0.710 & 0.047 \\
CAC      & 10 & 2.7  & 0.589 & 0.041 & 9  & 1.5  & 0.504 & 0.038 & 9  & 2.3  & 0.550 & 0.040 \\
DAX      & 11 & 2.0  & 0.661 & 0.044 & 9  & 1.2  & 0.539 & 0.038 & 12 & 4.1  & 0.738 & 0.051 \\
VIX      & 13 & 1.8  & 0.803 & 0.050 & 11 & 0.5  & 0.721 & 0.044 & 11 & 1.9  & 0.726 & 0.046 \\
EURUSD   & 8  & 1.1  & 0.471 & 0.034 & 9  & 0.9  & 0.562 & 0.037 & 9  & 1.8  & 0.571 & 0.040 \\
EURGBP   & 12 & 2.7  & 0.707 & 0.047 & 14 & 2.0  & 0.866 & 0.054 & 10 & 2.4  & 0.631 & 0.043 \\
EURCHF   & 15 & 4.5  & 0.874 & 0.057 & 14 & 4.9  & 0.823 & 0.055 & 16 & 7.6  & 0.984 & 0.065 \\
EURCNY   & 14 & 3.4  & 0.830 & 0.054 & 14 & 2.1  & 0.868 & 0.054 & 10 & 1.5  & 0.654 & 0.043 \\
EURRUB   & 11 & 1.3  & 0.679 & 0.044 & 10 & 0.8  & 0.626 & 0.041 & 13 & 3.9  & 0.832 & 0.054 \\
ECO      & 14 & 3.5  & 0.832 & 0.054 & 11 & 1.1  & 0.706 & 0.044 & 14 & 5.3  & 0.881 & 0.058 \\
LP01TREU & 17 & 7.6  & 0.974 & 0.064 & 16 & 7.6  & 0.926 & 0.062 & 15 & 6.5  & 0.931 & 0.061 \\
XAU      & 12 & 1.9  & 0.734 & 0.047 & 13 & 2.1  & 0.793 & 0.051 & 13 & 3.6  & 0.836 & 0.054 \\
LBEATREU & 13 & 1.9  & 0.805 & 0.050 & 13 & 2.7  & 0.805 & 0.051 & 11 & 1.7  & 0.744 & 0.046 \\
\bottomrule
\end{tabular}
\end{table*}

\begin{table*}[htbp]
\centering
\caption{Node-level centrality measures for the negative extreme GGM across
sample periods.}\label{tab:app:centrality_neg}
\small
\begin{tabular}{lrrrrrrrrrrrr}
\toprule
 & \multicolumn{4}{c}{Full} & \multicolumn{4}{c}{Phase~3} &
   \multicolumn{4}{c}{Phase~4} \\
\cmidrule(lr){2-5}\cmidrule(lr){6-9}\cmidrule(lr){10-13}
Node & Deg & Betw & Eigen & PR & Deg & Betw & Eigen & PR & Deg & Betw & Eigen & PR \\
\midrule
MO1      & 18 & 8.6  & 1.000 & 0.069 & 16 & 8.5  & 0.945 & 0.063 & 13 & 5.1  & 0.910 & 0.058 \\
CO1      & 12 & 1.4  & 0.742 & 0.048 & 12 & 1.5  & 0.793 & 0.048 & 11 & 1.8  & 0.824 & 0.049 \\
XA1      & 15 & 3.2  & 0.893 & 0.058 & 16 & 4.6  & 1.000 & 0.062 & 12 & 4.0  & 0.842 & 0.054 \\
NG1      & 15 & 3.2  & 0.893 & 0.058 & 14 & 3.0  & 0.892 & 0.055 & 14 & 5.0  & 0.987 & 0.061 \\
SPX      & 16 & 9.2  & 0.880 & 0.063 & 17 & 11.6 & 0.984 & 0.067 & 15 & 12.4 & 0.946 & 0.066 \\
SXXP     & 6  & 0.1  & 0.320 & 0.028 & 6  & 0.0  & 0.349 & 0.029 & 6  & 0.9  & 0.340 & 0.031 \\
MXEU0EN  & 10 & 2.3  & 0.568 & 0.042 & 10 & 1.9  & 0.609 & 0.043 & 11 & 6.8  & 0.716 & 0.051 \\
SPGTCED  & 17 & 9.6  & 0.928 & 0.066 & 14 & 5.4  & 0.838 & 0.056 & 10 & 5.0  & 0.645 & 0.047 \\
CAC      & 9  & 0.9  & 0.513 & 0.039 & 8  & 0.6  & 0.477 & 0.036 & 8  & 2.0  & 0.494 & 0.039 \\
DAX      & 9  & 0.9  & 0.513 & 0.039 & 11 & 2.5  & 0.670 & 0.046 & 9  & 3.8  & 0.531 & 0.043 \\
VIX      & 12 & 0.9  & 0.747 & 0.048 & 11 & 0.7  & 0.744 & 0.045 & 11 & 1.7  & 0.833 & 0.049 \\
EURUSD   & 8  & 0.5  & 0.510 & 0.034 & 7  & 0.5  & 0.459 & 0.031 & 9  & 1.9  & 0.653 & 0.042 \\
EURGBP   & 14 & 2.7  & 0.835 & 0.055 & 14 & 3.2  & 0.874 & 0.056 & 13 & 3.8  & 0.920 & 0.057 \\
EURCHF   & 16 & 6.7  & 0.898 & 0.062 & 14 & 6.2  & 0.837 & 0.056 & 13 & 5.3  & 0.897 & 0.058 \\
EURCNY   & 15 & 5.3  & 0.857 & 0.059 & 14 & 3.7  & 0.871 & 0.056 & 11 & 4.3  & 0.770 & 0.050 \\
EURRUB   & 10 & 0.8  & 0.609 & 0.041 & 10 & 0.9  & 0.637 & 0.042 & 14 & 4.0  & 1.000 & 0.061 \\
ECO      & 9  & 0.2  & 0.590 & 0.037 & 10 & 0.6  & 0.679 & 0.041 & 10 & 2.4  & 0.718 & 0.046 \\
LP01TREU & 15 & 6.2  & 0.851 & 0.059 & 16 & 7.7  & 0.965 & 0.063 & 9  & 3.9  & 0.568 & 0.043 \\
XAU      & 13 & 2.6  & 0.767 & 0.051 & 13 & 2.4  & 0.822 & 0.052 & 10 & 3.0  & 0.692 & 0.046 \\
LBEATREU & 11 & 0.7  & 0.697 & 0.044 & 13 & 2.4  & 0.840 & 0.052 & 11 & 2.9  & 0.777 & 0.050 \\
\bottomrule
\end{tabular}
\end{table*}

\begin{table*}[htbp]
\centering
\caption{MO1 pairwise extremal dependence $\hat{\chi}$ with all other
variables, by tail direction and sample period. Partners are sorted from
strongest to weakest by their maximum value across the six
configurations.}\label{tab:app:chi}
\begin{tabular}{lrrrrrr}
\toprule
 & \multicolumn{2}{c}{Full} & \multicolumn{2}{c}{Phase~3} &
   \multicolumn{2}{c}{Phase~4} \\
\cmidrule(lr){2-3}\cmidrule(lr){4-5}\cmidrule(lr){6-7}
Partner & Pos & Neg & Pos & Neg & Pos & Neg \\
\midrule
XA1      & 0.28 & 0.30 & 0.30 & 0.28 & 0.28 & 0.35 \\
CO1      & 0.26 & 0.27 & 0.30 & 0.28 & 0.22 & 0.26 \\
MXEU0EN  & 0.26 & 0.29 & 0.28 & 0.29 & 0.24 & 0.27 \\
DAX      & 0.24 & 0.27 & 0.25 & 0.27 & 0.24 & 0.28 \\
SPX      & 0.23 & 0.27 & 0.26 & 0.28 & 0.19 & 0.25 \\
SXXP     & 0.24 & 0.27 & 0.26 & 0.27 & 0.21 & 0.27 \\
CAC      & 0.25 & 0.27 & 0.26 & 0.29 & 0.22 & 0.25 \\
ECO      & 0.22 & 0.26 & 0.24 & 0.27 & 0.20 & 0.26 \\
LP01TREU & 0.24 & 0.26 & 0.25 & 0.27 & 0.19 & 0.24 \\
EURRUB   & 0.20 & 0.19 & 0.18 & 0.19 & 0.25 & 0.18 \\
NG1      & 0.21 & 0.23 & 0.22 & 0.24 & 0.20 & 0.23 \\
XAU      & 0.18 & 0.24 & 0.18 & 0.24 & 0.17 & 0.23 \\
SPGTCED  & 0.23 & 0.25 & 0.24 & 0.27 & 0.22 & 0.23 \\
EURCNY   & 0.20 & 0.22 & 0.22 & 0.22 & 0.18 & 0.23 \\
EURUSD   & 0.20 & 0.23 & 0.20 & 0.23 & 0.21 & 0.21 \\
EURCHF   & 0.19 & 0.20 & 0.20 & 0.20 & 0.19 & 0.19 \\
EURGBP   & 0.19 & 0.17 & 0.18 & 0.18 & 0.20 & 0.16 \\
LBEATREU & 0.18 & 0.18 & 0.17 & 0.19 & 0.19 & 0.15 \\
VIX      & 0.17 & 0.16 & 0.16 & 0.17 & 0.18 & 0.14 \\
\bottomrule
\end{tabular}
\end{table*}

\bigskip

Tables~\ref{tab:app:ergm_m0}--\ref{tab:app:ergm_selection} report the complete
ERGM estimation results for the M0--M3 models across all nine networks,
together with AIC and BIC model selection statistics. The selected model
varies across networks but converges on M2 by AIC for six of the nine
configurations. The full-sample positive extreme network is degenerate from
M2 onwards (the AIC-preferred model for this single network is M1). The
Phase~4 standard GGM is best fit by M3 (gwdsp replacing gwesp). M3 is
degenerate for all six extreme networks in the full and Phase~3 samples.

\begin{table*}[htbp]
\centering
\caption{M0 ERGM (edges only) for all nine networks. Standard errors in
parentheses. Significance: $^{***}p<0.001$, $^{**}p<0.01$, $^{*}p<0.05$,
$^{.}p<0.1$.}\label{tab:app:ergm_m0}
\begin{tabular}{l rr rr rr}
\toprule
 & \multicolumn{2}{c}{Standard} & \multicolumn{2}{c}{Pos.\ extreme} &
   \multicolumn{2}{c}{Neg.\ extreme} \\
\cmidrule(lr){2-3}\cmidrule(lr){4-5}\cmidrule(lr){6-7}
Period & $\hat{\theta}$ & $\hat{p}$ & $\hat{\theta}$ & $\hat{p}$ &
         $\hat{\theta}$ & $\hat{p}$ \\
\midrule
Full     & $-$0.297$^{*}$ (0.147)   & 0.426
         & 0.725$^{***}$ (0.155)    & 0.674
         & 0.654$^{***}$ (0.153)    & 0.658 \\
Phase~3  & $-$0.631$^{***}$ (0.152) & 0.347
         & 0.701$^{***}$ (0.154)    & 0.668
         & 0.607$^{***}$ (0.152)    & 0.647 \\
Phase~4  & $-$0.798$^{***}$ (0.157) & 0.311
         & 0.516$^{***}$ (0.150)    & 0.626
         & 0.318$^{*}$ (0.147)      & 0.579 \\
\bottomrule
\end{tabular}
\end{table*}

\begin{table*}[htbp]
\centering
\caption{M1 ERGM (edges plus sector nodefactor) for all nine networks. Bond
is the reference category. Standard errors in parentheses. Significance as
in Table~\ref{tab:app:ergm_m0}.}\label{tab:app:ergm_m1}
\resizebox{\textwidth}{!}{
\begin{tabular}{l rrr rrr rrr}
\toprule
 & \multicolumn{3}{c}{Full} & \multicolumn{3}{c}{Phase~3} &
   \multicolumn{3}{c}{Phase~4} \\
\cmidrule(lr){2-4}\cmidrule(lr){5-7}\cmidrule(lr){8-10}
Term & Std & Pos & Neg & Std & Pos & Neg & Std & Pos & Neg \\
\midrule
edges
  & 0.11 (0.65)
  & 2.03$^{*}$ (0.81)
  & 0.92 (0.71)
  & 0.04 (0.66)
  & 1.76$^{*}$ (0.78)
  & 1.83$^{*}$ (0.78)
  & $-$0.75 (0.71)
  & 1.07 (0.71)
  & $-$0.14 (0.66) \\
nodefactor: Clean Eq.
  & $-$0.23 (0.48)
  & $-$0.46 (0.56)
  & $-$0.00 (0.52)
  & $-$0.12 (0.49)
  & $-$0.70 (0.54)
  & $-$0.68 (0.53)
  & 0.00 (0.52)
  & $-$0.13 (0.51)
  & 0.00 (0.48) \\
nodefactor: Commodity
  & $-$0.46 (0.42)
  & $-$0.00 (0.50)
  & 0.59 (0.47)
  & $-$0.91$^{*}$ (0.44)
  & 0.54 (0.51)
  & $-$0.00 (0.48)
  & $-$0.97$^{*}$ (0.49)
  & 0.13 (0.45)
  & 0.59 (0.42) \\
nodefactor: Equity
  & $-$0.04 (0.40)
  & $-$1.15$^{*}$ (0.47)
  & $-$0.74$^{.}$ (0.43)
  & $-$0.19 (0.41)
  & $-$1.10$^{*}$ (0.46)
  & $-$1.07$^{*}$ (0.46)
  & 0.26 (0.43)
  & $-$0.67 (0.42)
  & $-$0.05 (0.40) \\
nodefactor: FX
  & $-$0.14 (0.40)
  & $-$0.85$^{.}$ (0.47)
  & $-$0.11 (0.43)
  & $-$0.39 (0.41)
  & $-$0.65 (0.46)
  & $-$0.73 (0.45)
  & 0.36 (0.43)
  & $-$0.35 (0.42)
  & 0.47 (0.40) \\
nodefactor: Volatility
  & $-$0.46 (0.48)
  & $-$0.73 (0.55)
  & $-$0.13 (0.51)
  & $-$0.12 (0.49)
  & $-$0.70 (0.54)
  & $-$0.68 (0.53)
  & $-$0.28 (0.53)
  & $-$0.25 (0.50)
  & 0.11 (0.48) \\
\bottomrule
\end{tabular}}
\end{table*}

\begin{table*}[htbp]
\centering
\caption{Model selection statistics (AIC and BIC) for the M0--M2 ERGM
specifications across all nine networks, computed via MPLE. The model
preferred by each criterion is marked in bold.}\label{tab:app:ergm_selection}
\begin{tabular}{l rr rr rr}
\toprule
 & \multicolumn{2}{c}{M0} & \multicolumn{2}{c}{M1} &
   \multicolumn{2}{c}{M2} \\
\cmidrule(lr){2-3}\cmidrule(lr){4-5}\cmidrule(lr){6-7}
Network & AIC & BIC & AIC & BIC & AIC & BIC \\
\midrule
Full Standard
  & 260.64 & \textbf{263.89}
  & 267.68 & 287.16
  & \textbf{252.75} & 278.73 \\
Phase~3 Standard
  & 247.40 & 250.65
  & 250.54 & 270.02
  & \textbf{216.82} & \textbf{242.80} \\
Phase~4 Standard
  & 235.80 & \textbf{239.05}
  & 231.43 & 250.91
  & \textbf{228.64} & 254.61 \\
Full Pos.\ extreme
  & 241.98 & \textbf{245.23}
  & 237.88 & 257.36
  & \textbf{214.05} & 240.03 \\
Phase~3 Pos.\ extreme
  & 243.41 & 246.66
  & 231.66 & 251.14
  & \textbf{193.25} & \textbf{219.23} \\
Phase~4 Pos.\ extreme
  & 253.14 & \textbf{256.38}
  & 256.34 & 275.82
  & \textbf{230.51} & 256.49 \\
Full Neg.\ extreme
  & 246.12 & 249.37
  & 241.52 & 261.00
  & \textbf{179.01} & \textbf{204.98} \\
Phase~3 Neg.\ extreme
  & 248.64 & 251.89
  & 246.29 & 265.77
  & \textbf{192.49} & \textbf{218.47} \\
Phase~4 Neg.\ extreme
  & 260.64 & 263.89
  & 264.70 & 284.18
  & \textbf{237.64} & \textbf{263.62} \\
\bottomrule
\end{tabular}
\end{table*}

\end{document}